\documentclass[aps, pra, superscriptaddress, twocolumn]{revtex4}
\usepackage{hyperref}
\usepackage{graphicx}
\usepackage{amsmath}
\usepackage{amssymb}
\usepackage{amsfonts}   
\usepackage{multirow}
\usepackage[b]{esvect} 
\usepackage{braket}
\usepackage{color}

\newcommand{\Tr}{\mathrm{Tr}}

\newcommand{\bea}{\begin{eqnarray}}
\newcommand{\eea}{\end{eqnarray}}
\newcommand{\beq}{\begin{equation}}
\newcommand{\eeq}{\end{equation}}

\newcommand{\up}{\uparrow}
\newcommand{\down}{\downarrow}

\begin{document}
\title{Fermion pairing in mixed-dimensional atomic mixtures}

\author{Junichi Okamoto}
\affiliation{Zentrum f\"ur Optische Quantentechnologien and Institut f\"ur Laserphysik, Universit\"at Hamburg, 22761 Hamburg, Germany}
\affiliation{The Hamburg Centre for Ultrafast Imaging, Luruper Chaussee 149, 22761 Hamburg, Germany}

\author{Ludwig Mathey}
\affiliation{Zentrum f\"ur Optische Quantentechnologien and Institut f\"ur Laserphysik, Universit\"at Hamburg, 22761 Hamburg, Germany}
\affiliation{The Hamburg Centre for Ultrafast Imaging, Luruper Chaussee 149, 22761 Hamburg, Germany}

\author{Wen-Min Huang}
\email{wenmin@phys.nchu.edu.tw}
\affiliation{Department of Physics, National Chung-Hsing University, Taichung 40227, Taiwan}

\date{\today}

\begin{abstract}
We investigate the quantum phases of mixed-dimensional cold atom mixtures. In particular, we consider a mixture of a Fermi gas in a two-dimensional lattice, interacting with a bulk Fermi gas or a Bose-Einstein condensate in a three-dimensional lattice. The effective interaction of the two-dimensional system mediated by the bulk system is determined. We perform a functional renormalization group analysis, and demonstrate that by tuning the properties of the bulk system, a subtle competition of several superconducting orders can be controlled among $s$-wave, $p$-wave, $d_{x^2-y^2}$-wave, and $g_{xy(x^2-y^2)}$-wave pairing symmetries. Other instabilities such as a charge-density wave order are also demonstrated to occur. In particular, we find that the critical temperature of the $d$-wave pairing induced by the next-nearest-neighbor interactions can be an order of magnitude larger than that of the same pairing induced by doping in the simple Hubbard model. We expect that by combining the nearest-neighbor interaction with the next-nearest-neighbor hopping (known to enhance $d$-wave pairing), an even higher critical temperature may be achieved.

\end{abstract}

\maketitle
\section{introduction}
\begin{figure}[!bp]
\begin{center}
   \includegraphics[width = 0.65\columnwidth ]{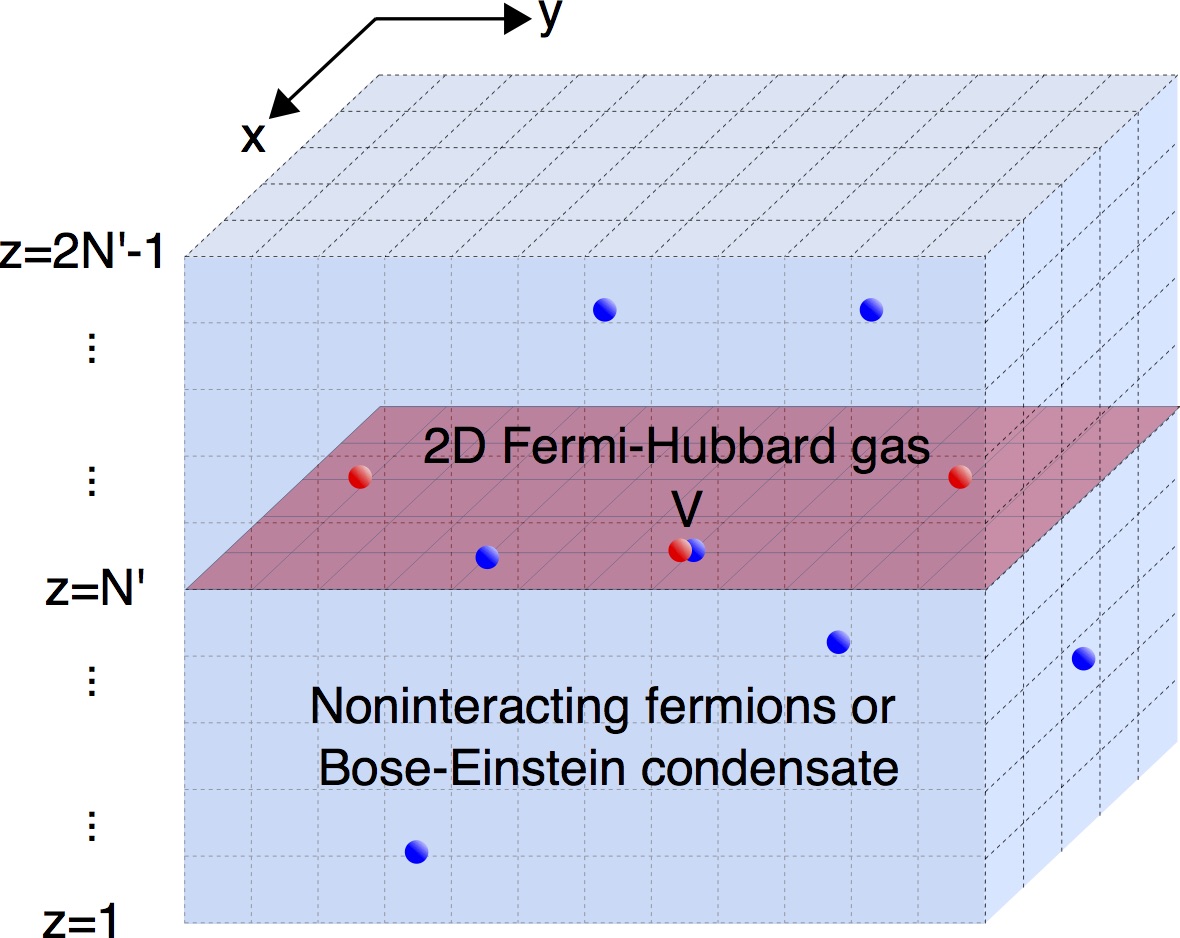}
   \caption{Schematics of the system that we consider. A 2D Fermi gas (red dots) interacts with 3D particles (blue dots) via a local density-density interaction $V$ at $z=N'$. We use an open boundary condition for the $z$ direction of the 3D system with $2N'-1$ sites.}
\label{system}
\end{center}
\end{figure}

Recent experiments of ultracold atoms in optical lattices successfully simulate various quantum lattice models, demonstrating quantitative agreement with theoretical predictions \cite{Lewenstein2007, Bloch2008}. One of the extensively studied models is the Hubbard model of fermions \cite{Kohl2005, Esslinger2010} and of bosons \cite{Jaksch1998, Greiner2002, Jaksch2005}, which has a local on-site interaction $U$. Until recently, only the on-site interaction has been realized in experiments, except long range interactions engineered in polar gases \cite{Lahaye2009, Dutta2015} or Rydberg dressed gases \cite{Johnson2010,Pupillo2010,Henkel2010,Honer2010, Balewski2014}. In this paper, we propose a way to generate an effective nonlocal interaction in a two-dimensional (2D) interacting Fermi gas by bringing it in contact with a three-dimensional (3D) system: either a noninteracting Fermi gas or a Bose-Einstein condensate (BEC) (Fig.~\ref{system}). The induced interaction creates a tunable 2D extended Fermi-Hubbard model \cite{Scalapino2012, Huang2013a} of ultracold atoms.


Experimentally mixed-dimensional atomic mixtures are realized by using species-specific optical lattices \cite{LeBlanc2007, Catani2009, Lamporesi2010, Haller2010}. Each atomic species is trapped by different optical potentials and confined into different geometries and dimensions. In such systems, confinement-induced resonances give rise to exotic pairing phenomena such as $p$-wave resonances and Efimov effects. These experiments have triggered numbers of theoretical investigations: Fermi-Fermi mixtures are studied in Refs.~\onlinecite{Nishida2008, Nishida2010a, Iskin2010, Yang2011, Huang2013, Kim2013}, Bose-Bose mixtures in Refs.~\onlinecite{Young2010, Minardi2011}, and Bose-Fermi mixtures in Refs.~\onlinecite{Malatsetxebarria2013, Malatsetxebarria2013a, Wu2016, Midtgaard2016, Caracanhas2017}. We note that most previous theoretical calculations in mixed-dimensional systems (except Refs.~\onlinecite{Huang2013, Malatsetxebarria2013}) are based on mean-field analysis, which is inadequate to give precise phase diagrams when the embedded system is one or two-dimensional. In order to incorporate strong quantum fluctuations in low dimensions, we use the functional renormalization group (fRG)\cite{Shankar1994, Halboth2000, Zanchi2000, Honerkamp2001, Honerkamp2001a, Kopietz2010introduction, Metzner2012, Platt2013}, which treats various competing orders without bias.

In this work, we focus on a 2D Fermi gas described as the Hubbard model, 
\begin{equation}
H_\text{2D} = -t  \sum_{\langle \mathbf{r} , \mathbf{r}' \rangle ,s} c_{\mathbf{r}s}^{\dagger} c_{\mathbf{r}' s} + U \sum_{\mathbf{r}} n_{\mathbf{r} \up} n_{\mathbf{r} \down} - \mu \sum_{\mathbf{r} ,s}  n_{\mathbf{r} s} ,
\end{equation}
where $c^{(\dagger)}_{\mathbf{r}s}$ is the annihilation (creation) operator of a fermion with spin $s$ at site $\mathbf{r} = (x,y)$, and $t$ is the hopping between nearest-neighbor (NN) sites $\langle \mathbf{r}, \mathbf{r}' \rangle$. $U$ is the on-site Hubbard interaction between particle densities $n_{\mathbf{r} s} = c^{\dagger}_{\mathbf{r} s}c_{\mathbf{r}s}$, and $\mu$ is the chemical potential. In the following, we assume that a 3D system, either a noninteracting Fermi gas or a BEC, is in contact with 2D fermions via a local density-density interaction $V$ at their interface (Fig.~\ref{system}). After integrating out the 3D degrees of freedom (ignoring the retardation effects), nonlocal effective interactions are left among the 2D fermions. A noninteracting 3D Fermi gas induces Ruderman-Kittel-Kasuya-Yosida-type interactions \cite{Ruderman1954, Kasuya1956, Yosida1957}; the effective interaction oscillates in space with the modulation period determined by the Fermi momentum of the bulk system. Thus, while the induced on-site interaction is always attractive, the NN interaction changes signs as the filling of the 3D system changes. When a BEC is in contact with the 2D fermions, induced on-site and NN interactions are always attractive \cite{Illuminati2004, Mathey2006, Mathey2007}. With these effective interactions as well as the original on-site Hubbard interaction $U$, we obtain zero-temperature phase diagrams by fRG. Various superconducting instabilities appear depending on the model parameters. In particular, we find that the gap of the $d$-wave pairing induced by the attractive NN interaction is larger than that of the pairing induced by doping in the simple Hubbard model.

The rest of this paper is organized as follows. Section \ref{Fermi-Fermi} considers the case of a Fermi-Fermi mixture. Section \ref{Fermi-BEC} treats the case of a Fermi-BEC mixture. Section \ref{discussion} discusses the effects of finite temperatures and the experimental relevance of our analysis. Section \ref{conclusion} is a conclusion. The Appendix considers the 2D limit of the Fermi-Fermi mixture.

\section{Fermi-Fermi mixture}
\label{Fermi-Fermi}
\subsection{Model}
First, we study the case when the 3D system is a noninteracting Fermi gas. We assume a single component gas, which is easier to realize in experiments. The Hamiltonian of the 3D part is 
\begin{equation}
H_\text{3D}^\text{F} = -t'  \sum_{\langle \mathbf{r}z , \mathbf{r}' z' \rangle } d_{\mathbf{r}z }^{\dagger} d_{\mathbf{r}' z'} - \mu' \sum_{\mathbf{r}, z } d_{\mathbf{r} z}^{\dagger} d_{\mathbf{r} z },
\end{equation}
where $d^{(\dagger)}_{\mathbf{r} z}$ is the annihilation (creation) operator of a fermion at site $(\mathbf{r}, z) = (x,y,z)$, $t'$ is the hopping amplitude between NN sites $\langle \mathbf{r}z , \mathbf{r}' z' \rangle$, and $\mu'$ is the chemical potential. We use periodic boundary conditions for $x$ and $y$ directions with $N\times N$ sites, while we assume an open boundary condition with $2N'-1$ sites along the $z$ axis. In momentum space, we obtain
\begin{equation}
H_\text{3D}^\text{F} = \sum_{\mathbf{p} ,k} \xi_{\mathbf{p}k} d_{\mathbf{p} k}^{\dagger} d_{\mathbf{p} k} ,
\end{equation}
with a 3D dispersion 
\begin{equation}
\xi_{\mathbf{p} k} = -2 t' \left[ \cos\left( p_x a \right)+\cos\left(p_y a \right) + \cos (k a ) \right]- \mu' .
\end{equation}
Here $a$ is the lattice constant, $\mathbf{p} = (p_x, p_y) = \left( n_x, n_y \right)2 \pi /N a$ ($n_{x,y} \in \mathbb{Z}$ and $1\leq n_{x,y} \leq N$) is the in-plane momentum, and $k = \pi n_z/2N'a$ ($n_z \in \mathbb{Z}$ and $1 \leq n_z \leq 2N'-1$) is the out-of-plane momentum along the $z$ axis, satisfying open boundary conditions. 

The local contact interaction between 2D and 3D fermions at $z=N'$ is 
\begin{equation}
H^\text{F}_\text{int} =  V \sum_{\mathbf{r} ,s}  n_{\mathbf{r}s} d_{\mathbf{r} N'}^{\dagger} d_{\mathbf{r}N'} .
\end{equation}

\subsection{Effective interaction}
In this section, we derive the effective interaction mediated by the bulk fermions. We integrate out the 3D fermions in a path-integral form. Such a procedure is legitimate if the 3D fermions are Fermi liquids. This requires the interspecies interaction $V$ to be small compared to the bandwidth $12t'$. Once the 3D fermions become highly anisotropic (approaching to the 2D limit), even a weak perturbation can trigger an instability, and  treating both species by fRG is a more appropriate procedure as in Ref.~\onlinecite{Lai2014}.

\begin{figure}[!tb]
\begin{center}
   \includegraphics[width = 0.9\columnwidth ]{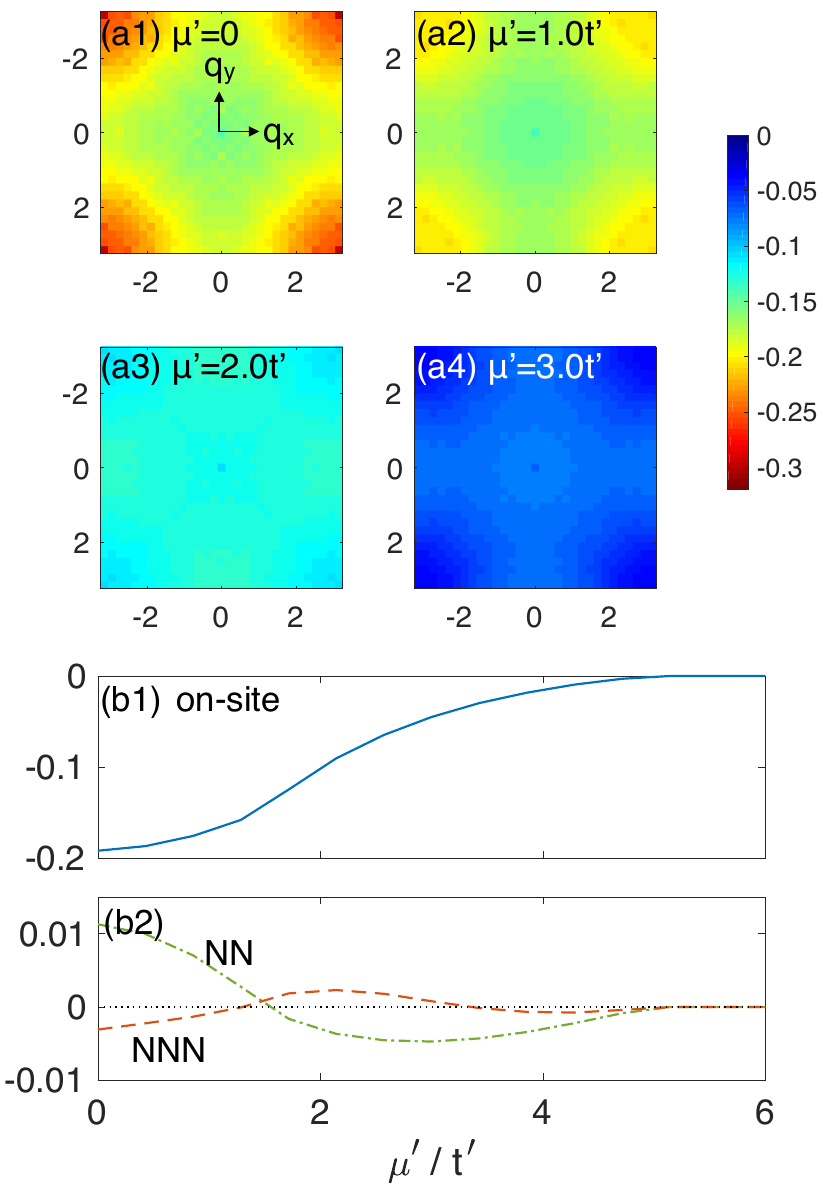}
   \caption{(a) The effective interaction given by Eq.~\eqref{UE_eff} in momentum space in units of $4V^2/t'$. We use $N=50$ and $N' = 11$. (a1)-(a4) represent the dependence of $U^\text{F}_\text{eff} (\mathbf{q})$ on $\mu'$. We observe a peak at $\mathbf{q} = (\pi,\pi)$ for $\mu'=0$ shifting to $(0,0)$ as $\mu'$ increases. (b) The effective interactions in real space in units of $4V^2/t'$. The NN and next-nearest-neighbor (NNN) interactions change signs as $\mu'$ changes. }
\label{Ueff}
\end{center}
\end{figure}

We start from the total action of the 2D and 3D fermions with the imaginary time $\tau$ and the inverse temperature $\beta$, 
\begin{multline}
S  = \int_0^{\beta} d\tau \Big[  \sum_{\mathbf{p}, s } c^{\dagger}_{\mathbf{p}  s} (\tau ) \partial_{\tau}c _{\mathbf{p}  s} (\tau )   + H_\text{2D}(\tau )  \\
+ \sum_{\mathbf{p} ,k } d^{\dagger}_{\mathbf{p}k } (\tau ) \partial_{\tau}d_{\mathbf{p}k  } (\tau ) + H^\text{F}_\text{3D}(\tau )  + H^\text{F}_\text{int}(\tau )  \Big].
\label{action}
\end{multline}
Introducing the Fourier series in Matsubara frequencies $\omega_n = \pi (2n+1)/\beta$ ($n \in \mathbb{Z}$) as 
\begin{equation}
c_{\mathbf{p} s} (\tau) = \frac{1}{\sqrt{\beta}} \sum_{n} e^{-i\omega_n \tau} c_{n \mathbf{p} s},
\end{equation}
we can write the quadratic action related to the 3D fermions [the second line in Eq.~\eqref{action}] in a matrix form,
\begin{equation}
S^{\text{F}}_\text{3D} =\sum_{n, n', \mathbf{p},\mathbf{p}', k ,k' }  d^{\dagger}_{n \mathbf{p} k }  \left[- G_0^{-1} + M \right]_{(n \mathbf{p} k);(n' \mathbf{p}' k')} d_{n' \mathbf{p}' k'} ,
\end{equation}
with a Green's function matrix $G_0$ and an interaction part $M$,
\begin{align}
[G_0]_{(n \mathbf{p} k);(n' \mathbf{p}' k')} &= \frac{1}{ i \omega_n - \xi_{\mathbf{p} k }} \delta_{nn'} \delta_{\mathbf{p} \mathbf{p}'} \delta_{k k'} ,\\
\begin{split}
[M]_{(n \mathbf{p} k);(n' \mathbf{p}' k')}  &= \frac{V}{N^2 N'\beta }\sin (kN'a) \sin(k' N' a) \\
&\times \sum_{m, \mathbf{q}, s}  c_{m \mathbf{q} s}^{\dagger} c_{n -n' +m, \mathbf{p} - \mathbf{p}' + \mathbf{q}, s}.
\end{split}
\end{align}
After integrating out the 3D fermions, the effective action for 2D fermions becomes
\begin{equation}
\begin{split}
S_\text{eff} &= S_\text{2D} -  \Tr \ln \left[ -G_0^{-1} + M \right] \\
&= S_\text{2D} - \Tr \ln  \left[ -G_0^{-1}(1 - G_0 M) \right] \\
&= S_\text{2D} + \sum_{n=1}^{\infty} \frac{ \Tr\left[ (G_0 M)^n \right]}{n} + \text{const}.
\end{split}
\end{equation}
We will ignore the self-energy correction to the 2D fermions corresponding to the first order in the expansion. The second-order term generates the effective interaction, 
\begin{multline}
\Tr\left[ (G_0 M)^2 \right] = \frac{V^2   }{N^4 N'^2 \beta^2 }  \sum_{n, n', \mathbf{p},\mathbf{p}', k ,k' }\\
 \Bigg[ \frac{\sin^2 (k N'a)}{ i \omega_n - \xi_{\mathbf{p} k }}  \sum_{m, \mathbf{q}, s}   c_{m \mathbf{q} s}^{\dagger} c_{n -n' +m, \mathbf{p} - \mathbf{p}' + \mathbf{q}, s}\\
\times \frac{\sin ^2(k' N'a)}{ i \omega_{n'} - \xi_{\mathbf{p}' k' }} \sum_{m',  \mathbf{q}', s'}  c_{m'  \mathbf{q}' s'}^{\dagger} c_{n' -n +m', \mathbf{p}' - \mathbf{p} + \mathbf{q}', s'} \Bigg] .
\end{multline}
Summation over $\omega_n$ gives the particle-hole propagator of the 3D fermions as
\begin{multline}
\Tr\left[ (G_0 M)^2 \right] =  \frac{V^2   }{N^4 N'^2 \beta } \sum_{\tilde{l}, \mathbf{p},\mathbf{p}', k ,k' } \\\Bigg[  \frac{n_\text{F}(\xi_{\mathbf{p} k }) - n_\text{F}(\xi_{\mathbf{p}+\mathbf{p}',  k' })}{i \tilde{\omega}_l + \xi_{\mathbf{p} k } - \xi_{\mathbf{p} + \mathbf{p}' ,k' }}  \sin^2 (kN'a ) \sin^2 (k' N'a )   \\
\times \sum_{m,m', \mathbf{q}, \mathbf{q}', s', s'} c_{m \mathbf{q} s}^{\dagger} c_{m-\tilde{l}, - \mathbf{p}'  + \mathbf{q}, s} c_{m'  \mathbf{q}' s'}^{\dagger} c_{m' + \tilde{l}, \mathbf{p}'  + \mathbf{q}', s'} \Bigg] ,
\end{multline}
where $\tilde{\omega}_l = \omega_{n'} - \omega_n$ is a bosonic Matsubara frequency, and $n_\text{F}(\xi)$ is the Fermi distribution function. If we consider $t' \gg t$, we can ignore the retardation effects, i.e., only the $\tilde{\omega}_l = 0$ component is important. In this limit, we obtain effective interactions among 2D fermions as
\begin{equation}
H^\text{F}_\text{eff} =  \frac{1}{2 N^2 }\sum_{\mathbf{r},\mathbf{r}', \mathbf{q} ,s , s'} U^\text{F}_\text{eff} (\mathbf{q}) e^{i (\mathbf{r}'- \mathbf{r}) \mathbf{q}} 
   c_{ \mathbf{r} s}^{\dagger} c_{ \mathbf{r}' s'}^{\dagger} c_{ \mathbf{r}' s'}  c_{\mathbf{r}  s},
\label{HF_eff}
\end{equation}
with 
\begin{multline}
U^\text{F}_\text{eff} (\mathbf{q}) =
V^2  \sum_{\mathbf{p}, k ,k' } \frac{\sin^2 (k N' a) \sin^2(k' N' a) }{N^2 N'^2} \\
\times   \frac{n_\text{F}(\xi_{\mathbf{p} k }) - n_\text{F}(\xi_{\mathbf{p}+\mathbf{q},  k' })}{ \xi_{\mathbf{p} k } - \xi_{\mathbf{p} + \mathbf{q} ,k' }}.
 \label{UE_eff}
\end{multline}
We assume zero-temperature in the following calculations. As we see in Sec. \ref{discussion}, the critical temperatures of density-wave or pairing instabilities are smaller than $t$. Therefore, to observe these phases, the 3D fermions need to be also as cold as $t$ at least. Considering the assumption $t' \gg t$, using the zero-temperature propagator is reasonable and self-consistent in the regime of interests.

\begin{figure}[!tb]
\begin{center}
   \includegraphics[width = 0.95\columnwidth ]{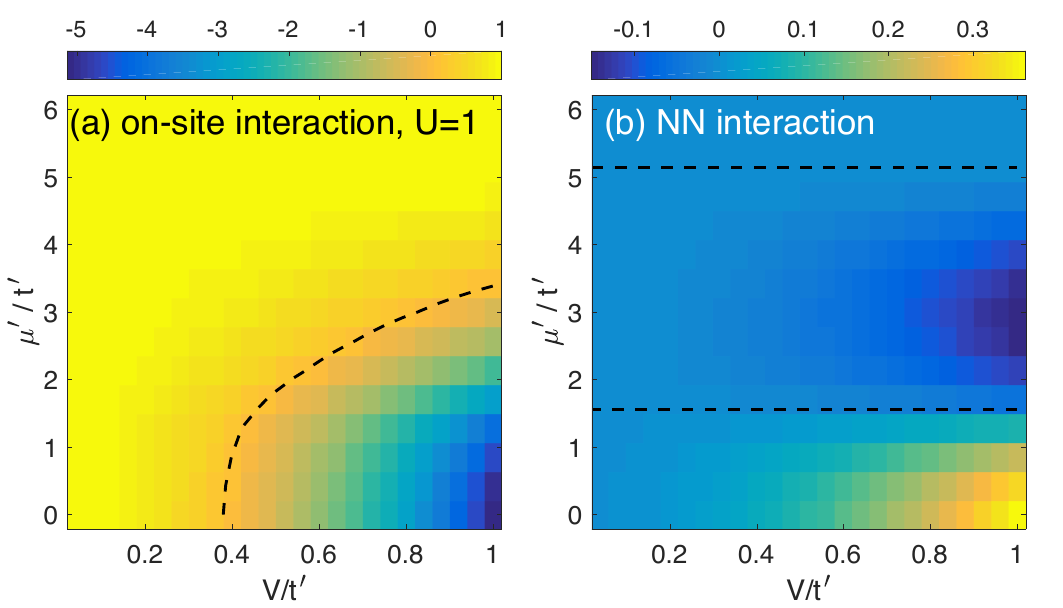}
   \caption{The total ``bare" on-site and nearest-neighbor interaction for 2D fermions with the original on-site interaction $U=1$. Dashed lines show the line where the interaction is zero.}
\label{U0_U1}
\end{center}
\end{figure}

We show the induced interaction in momentum space in Fig.~\ref{Ueff}(a) for $N'=11$. We note that the dependence on the bulk  thickness is marginal even in the limit of $N'=1$ as we discuss in the Appendix. The 3D fermionic bath induces an attractive interaction peaked at momentum $\mathbf{q}^*$ determined by the filling; $\mathbf{q}^* = (\pi,\pi)$ when $\mu' = 0$ at half-filling while it gradually shifts to $(0,0)$ as $\mu'$ increases.

In Fig.~\ref{Ueff}(b), we show the effective on-site, NN, and next-nearest-neighbor (NNN) interactions as functions of $\mu'$. In real space, the interaction oscillates with roughly the period $2\pi/\mathbf{q}^*$, while it also decays rapidly so that the NNN interaction is negligible. This indicates that we can map the 2D part of the system onto the extended Hubbard model with various interaction strengths by controlling the filling of the 3D fermions. We note that for the induced interaction to be non-negligible, the interspecies interaction $V$ needs to be larger than $t$. This is because the effective interaction is proportional to $V^2 / t'$, and we assume $t' \gg t$ to ignore the retardation effect. At the same time, to ignore the effect of $V$ on the 3D fermions, we need $V \ll 12t'$. 
 
The total on-site interaction, $U + U^\text{F}_\text{eff} (\mathbf{r}=0)$, and the NN interaction are plotted in Fig.~\ref{U0_U1} with $U=1$ for different $V$'s and $\mu'$'s. We can control these two interactions by shifting $U$, $V$, and $\mu'$. In Sec.~\ref{phase_diagram}, we show that the obtained phase diagrams can be well understood based on these values.

\subsection{Method}
\begin{figure}[!bt]
\begin{center}
   \includegraphics[width = 0.95\columnwidth ]{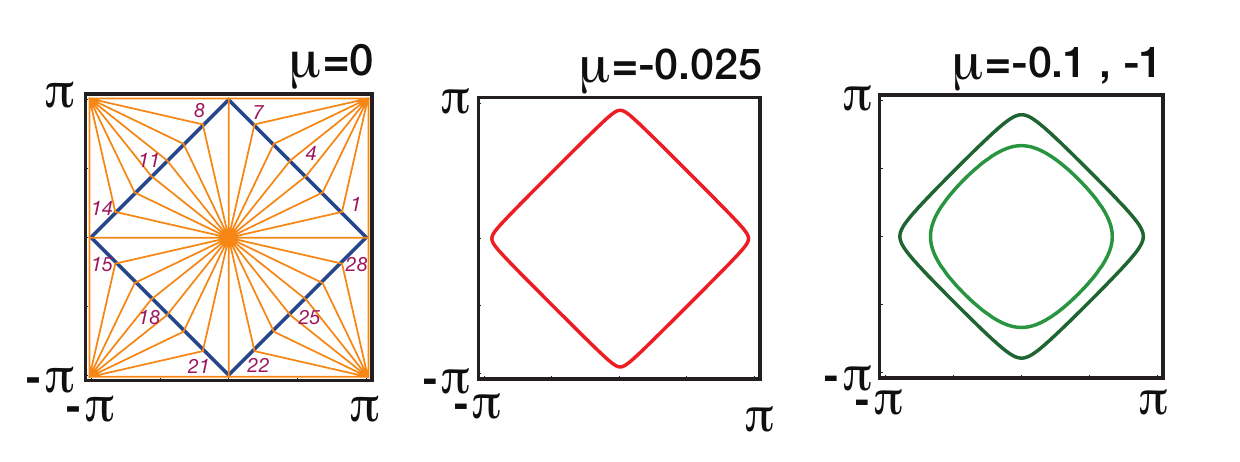}
   \caption{Schematics of the patching scheme we use and the Fermi surface. Thick lines represent Fermi surfaces for various chemical potentials in each panel [In the right panel, the inner (outer) path is for $\mu = -1 \ (-0.1)$.] Thin lines in the left panel show how to dissect the first Brillouin zone into narrow 28 patches. }
\label{BZ}
\end{center}
\end{figure}

Based on the effective interaction, phase diagrams are obtained by a fRG scheme. Here we briefly outline the standard $N$-patch scheme \cite{Shankar1994, Halboth2000, Zanchi2000, Honerkamp2001, Honerkamp2001a, Kopietz2010introduction, Metzner2012, Platt2013}, which we employ in this work. We divide the Brillouin zone into $N_\text{patch} = 28$ patches as shown in Fig.~\ref{BZ}. The $n$th patch has a patch momentum $\bar{\mathbf{k}}_n$ at the center of the Fermi surface. The interaction is now approximated as $U (\mathbf{k}_1, \mathbf{k}_2, \mathbf{k}_3) \rightarrow   U_{n_1 n_2 n_3}$, where $n_i$ is the patch that $\mathbf{k}_i$ belongs to, and the fourth momentum (not explicitly written above) is automatically determined by the momentum conservation. Naively the total number of coupling constants is $N_\text{patch}^3 = 21952$. However, we can reduce this number by using the symmetry of the Hamiltonian, and by ignoring the coupling constants that deviate from the momentum conservation significantly. The RG equation is obtained after integrating out the high-energy degrees of freedom around the ultraviolet cutoff $\Lambda$. By parametrizing the cutoff as $\Lambda(l) = \Lambda_0 e^{-l}$ with the initial value of the cutoff $\Lambda_0$, the coupling constants at lower energies are obtained by integrating the RG equations \cite{Kopietz2010introduction, Platt2013}:
\begin{widetext}
\begin{equation}
\begin{split}
\frac{\partial  U_{n_1 n_2 n_3}}{\partial l} &= - \sum_{n} \dot{\Pi}^- (n, \mathbf{q}_\text{pp}) \left( U_{n_1n_2n}U_{n_4 n_3 n} + U_{n_2 n_1 n}U_{n_3 n_4 n} \right)\\
 &+\sum_{n} \dot{\Pi}^+ (n, \mathbf{q}_\text{fs})  \left( 2U_{n n_4 n_1}U_{n n_2 n_3} -U_{n_4 n n_1}U_{n n_2 n_3}- U_{n n_4 n_1}U_{n_2 n n_3}\right)\\
 &+\sum_{n} \dot{\Pi}^+ (n, -\mathbf{q}_\text{fs})  \left( 2U_{n n_1 n_4}U_{n n_3 n_2} -U_{n_1 n n_4}U_{n n_3 n_2}- U_{n n_1 n_4}U_{n_3 n n_2}\right)\\
&- \sum_{n} \dot{\Pi}^+ (n, \mathbf{q}_\text{ex}) U_{n_3 n n_1}U_{n_2 n n_4} - \sum_{n} \dot{\Pi}^+ (n, -\mathbf{q}_\text{ex}) U_{n_1 n n_3}U_{n_4 n n_2},
\label{RG_Eq}
\end{split}
\end{equation}
\end{widetext}
where $\mathbf{q}_\text{pp}=\bar{\mathbf{k}}_{n_1}+\bar{\mathbf{k}}_{n_2}$, $\mathbf{q}_\text{fs}=\bar{\mathbf{k}}_{n_3}-\bar{\mathbf{k}}_{n_2}$, and $\mathbf{q}_\text{ex}=\bar{\mathbf{k}}_{n_1}-\bar{\mathbf{k}}_{n_3}$. $\dot{\Pi}^{\pm} (n, \mathbf{Q})$ is a differential of a bubble integral over frequency $\omega$ and momentum $\mathbf{k}$ inside the $n$th patch,
\begin{equation}
\dot{\Pi}^{\pm} (n, \mathbf{q}) = \pm \Lambda \int_{\omega} \int_{\mathbf{k} \in n} \dot{G}(\omega, \mathbf{k}) G[\pm \omega, \pm(\mathbf{k} - \mathbf{q})],
\end{equation}
with $G(\omega, \mathbf{k}) = \Theta(|\epsilon_\mathbf{k}|- \Lambda)/(i\omega-\epsilon_\mathbf{k})$ and $\epsilon_{\mathbf{k}} = -2t \left[\cos(k_x a) + \cos(k_y a) \right] - \mu$. Use of this free propagator means that we ignore the self-energy correction along the RG flows. 

A RG flow is started from an ultraviolet cutoff $\Lambda_0 \simeq 4t$ and integrated until one of the coupling constant becomes $\sim 30t$ or $\Lambda = 10^{-6} t$. The former indicates an ordering instability, while the latter indicates no instability, i.e., the Fermi liquid fixed point. To figure out the dominant instability, we decompose the renormalized interaction $\tilde{U}$ into six channels: spin-density wave (SDW), charge-density wave (CDW), ferromagnetic, Pomeranchuk, spin-singlet superconductivity (sSC), and spin-triplet superconductivity (tSC) orders \cite{Platt2013, Lai2014}, 
\begin{multline}
\sum_{\mathbf{k}_1, \mathbf{k}_2, \mathbf{k}_3, s, s'}\tilde{U} (\mathbf{k}_1, \mathbf{k}_2, \mathbf{k}_3) c^{\dagger}_{\mathbf{k}_1,s} c^{\dagger}_{\mathbf{k}_2,s} c_{\mathbf{k}_3,s'}  c_{-\mathbf{k}_1 - \mathbf{k}_2- \mathbf{k}_3,s} \\
= \sum_{i=\text{SDW}, \dots, \text{tSC}} \sum_{\mathbf{k}_1, \mathbf{k}_2} W^i (\mathbf{k}_1, \mathbf{k}_2) \mathcal{O}^{i, \dagger}_{\mathbf{k}_1} \mathcal{O}^i_{\mathbf{k}_2},
\end{multline}
where $\mathcal{O}^i_\mathbf{k}$ is the order parameter given by fermion bilinears. For example, the spin-singlet SC has  
\begin{align}
W^\text{sSC} (\mathbf{k}_1, \mathbf{k}_2) &= \tilde{U}(\mathbf{k}_1, -\mathbf{k}_1, -\mathbf{k}_2) + \tilde{U}(-\mathbf{k}_1, \mathbf{k}_1, -\mathbf{k}_2) ,\\ 
\mathcal{O}^\text{sSC}_\mathbf{k} &= \frac{1}{\sqrt{2}}  \left(c_{\mathbf{k} \up}c_{-\mathbf{k} \down} - c_{\mathbf{k} \down}c_{-\mathbf{k} \up} \right) .
\end{align}
With the patch approximation, $W^i (\mathbf{k}_1, \mathbf{k}_2)$ can be expressed as a $N_\text{patch}\times N_\text{patch}$ Hermitian matrix $\tilde{W}^i$ whose $(n_1, n_2)$ component is $\tilde{W}^i_{n_1  n_2} = W^i (\bar{\mathbf{k}}_{n_1}, \bar{\mathbf{k}}_{n_2})$. We diagonalize these matrices into the following forms
\begin{equation}
\tilde{W}^i_{n_1  n_2} = \sum_{\lambda = 1}^{N_\text{patch}} \omega^{i}_\lambda f^{i,*}_{\lambda} ( \bar{\mathbf{k}}_{n_1} ) f^{i}_{\lambda} ( \bar{\mathbf{k}}_{n_2} ).
\end{equation}
$f^{i}_{\lambda} ( \bar{\mathbf{k}}_{n} )$ gives the form factor of the order parameter. The leading instability is the one with the largest negative eigenvalue among $\omega^{i}_\lambda$'s. 
 
\subsection{Phase diagrams}
\label{phase_diagram}
\begin{figure*}[!tb]
\begin{center}
   \includegraphics[width = 0.9\textwidth ]{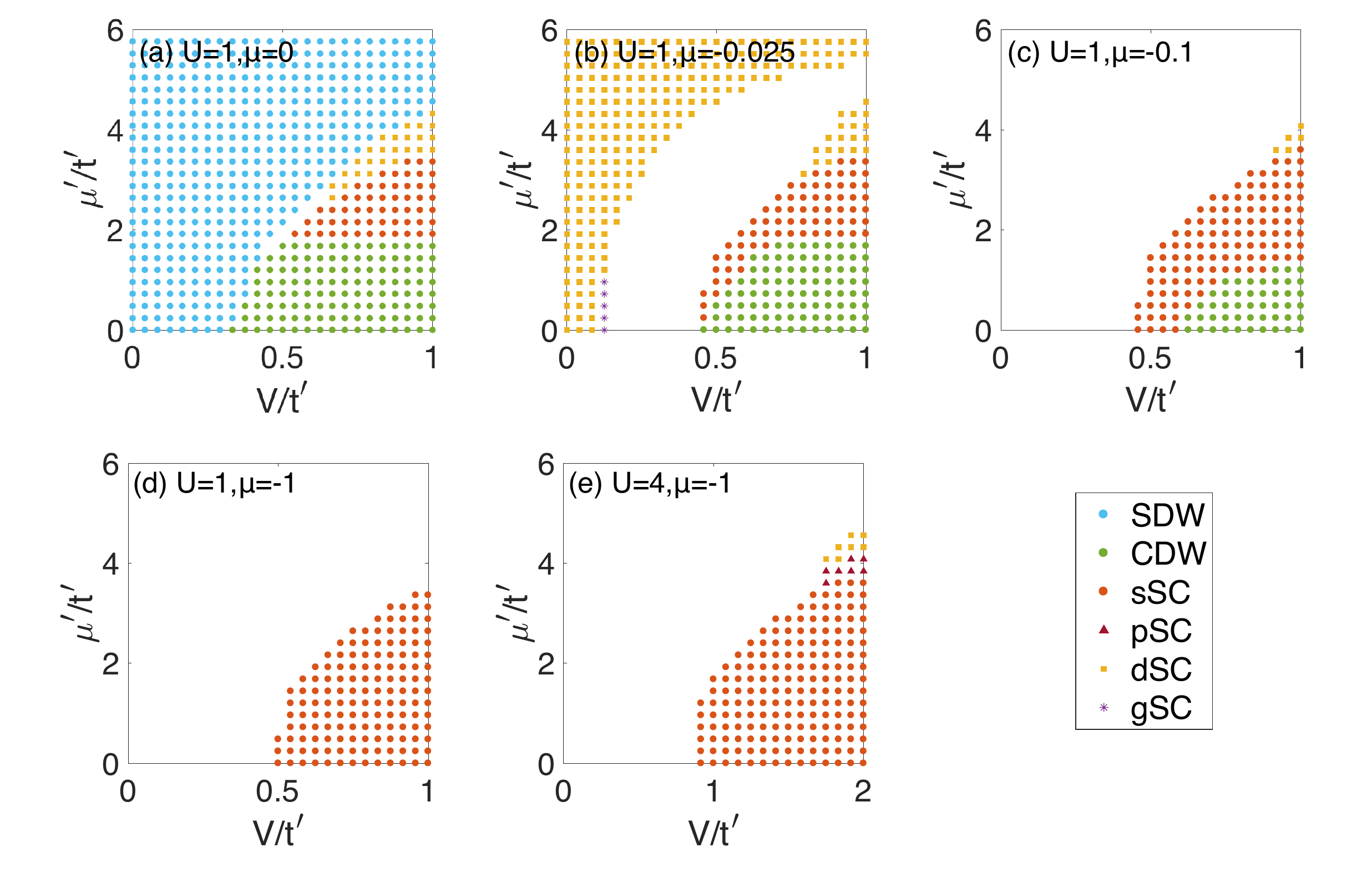}
   \caption{Phase diagrams of a 2D Fermi gas in contact with a noninteracting 3D Fermi gas. We use $N=50$, $N'=11$, and $t' = 8t$. Blank regions correspond to Fermi liquid; no instability is found.}
\label{PD_Fermi}
\end{center}
\end{figure*}
Fig.~\ref{PD_Fermi} shows phase diagrams as a function of $\mu'$ and $V$ for various cases. We use $N=50$, $N'=11$, and $t' = 8t$. At half-filling ($\mu=0$) [Fig.~\ref{PD_Fermi}(a)], we found SDW, CDW, $s$-wave SC, and $d_{x^2-y^2}$-wave SC. These can be well understood from the effective interactions in Fig.~\ref{U0_U1}. When $V$ is weak, the original repulsive Hubbard interaction $U$ is dominant, which leads to SDW. At small $\mu'$, once $V$ is strong, the on-site interaction becomes attractive, while the NN interaction is repulsive. This naturally leads to CDW with an ordering vector at $(\pi, \pi)$. At intermediate $\mu' $, the NN interaction becomes attractive as well as the on-site one, leading to $s$-wave SC and $d_{x^2-y^2}$-wave SC. We would like to emphasize that this $d_{x^2-y^2}$-wave SC is induced by the attractive NN interaction. At large $\mu'$, the induced interactions are weak, and the phase goes back to SDW dominated by the original on-site $U$. 

When the 2D system is slightly away from the half-filling $\mu = - 0.025$ [Fig.~\ref{PD_Fermi}(b)], SDW is replaced by $d_{x^2-y^2}$-wave SC, as the simple Hubbard model \cite{Halboth2000, Zanchi2000, Honerkamp2001}. We note that this $d$-SC is induced by doping, and can be found even at $V=0$. We also find $g_{xy(x^2-y^2)}$-wave SC at $V=t'/8$ and low $\mu'$. However, this instability is relevant only at very low energy scales $\Lambda \sim 2 \times10^{-6} t$, and therefore may not be experimentally realizable. As the filling is further reduced, the Fermi-liquid fixed point becomes dominant for most of the parameter regions [Figs.~\ref{PD_Fermi}(c) and (d)], while still $s$-wave pairing is induced for strong $V$ and small $\mu'$. We note that mapping the induced interactions onto on-site and NN interactions allows us to explore a wide range of parameters of the extended Hubbard model. For example, if we take $U=4t$, we can create $p$-wave SC, which was found in Ref.~\onlinecite{Huang2013a} [Fig.~\ref{PD_Fermi}(e)].

\section{Fermi-BEC mixture}
\label{Fermi-BEC}
\subsection{Model}
Next, we turn to the case when the 3D system is a BEC. Assuming weakly interacting bosons, the Hamiltonian is 
\begin{multline}
H_\text{3D}^{\text{B}} = -t_\text{B}  \sum_{\langle \mathbf{r}z , \mathbf{r}' z' \rangle} \psi_{\mathbf{r}z}^{\dagger}  \psi_{\mathbf{r}' z' }  - \mu_\text{B} \sum_{\mathbf{r}, z}  \psi_{\mathbf{r} z }^{\dagger}\psi_{\mathbf{r} z }  \\
+  \frac{U_\text{B}}{2}  \sum_{\mathbf{r}, z }  \psi_{\mathbf{r} z }^{\dagger}  \psi_{\mathbf{r} z }^{\dagger}\psi_{\mathbf{r} z }\psi_{\mathbf{r} z },
\end{multline}
where $\psi_{\mathbf{r}z}^{(\dagger)}$ is the annihilation (creation) operator of a boson at site $(\mathbf{r}, z)$, $t_\text{B}$ is the hopping amplitude, $\mu_\text{B}$ is the chemical potential, and $U_\text{B}$ is the weak on-site interaction. We decompose the wave function into the condensed part $\eta_{z}$ and fluctuations $\phi_{\mathbf{r}z}$ \cite{pethick2002bose, lewenstein2012ultracold},
\begin{equation}
\psi_{\mathbf{r}z} = \eta_{z}   + \phi_{\mathbf{r}z},
\end{equation}
where we assume a uniform condensate along the in-plane directions (along the $z$ axis, the open boundary condition induces modulation). Up to quadratic order in the fluctuations, this leads to
\begin{multline}
H_\text{3D}^{\text{B}} \simeq -t_\text{B}  \sum_{\langle \mathbf{r}z , \mathbf{r}' z' \rangle } \phi_{\mathbf{r}z}^{\dagger}  \phi_{\mathbf{r}' z' } - \mu_\text{B} \sum_{\mathbf{r}, z }  \phi_{\mathbf{r} z }^{\dagger}\phi_{\mathbf{r} z } \\
+ \frac{U_\text{B}}{2}   \sum_{\mathbf{r}, z } \eta_z^2 \left(4 \phi_{\mathbf{r} z }^{\dagger}\phi_{\mathbf{r} z } + \phi_{\mathbf{r} z }^{\dagger}\phi_{\mathbf{r} z }^{\dagger} + \phi_{\mathbf{r} z }\phi_{\mathbf{r} z } \right) ,
\label{HB_eff}
\end{multline}
if the condensed part satisfies the Gross-Pitaevskii equation
\begin{equation}
-t_\text{B}  \left[\sum_{z'} (\delta_{z',z+1}+\delta_{z',z-1})\eta_{z'} \right]-\mu_\text{B} \eta_z + U_\text{B} \eta_{z}^3 =0.
\label{GPE}
\end{equation}
The interaction between 2D fermions and a BEC is approximated as 
\begin{equation}
\begin{split}
H_\text{int}^\text{B} &=  V \sum_{\mathbf{r} ,s} n_{\mathbf{r} s} \psi_{\mathbf{r} N'}^{\dagger} \psi_{\mathbf{r}N' } \\
&\simeq V \sum_{\mathbf{r} ,s} n_{\mathbf{r} s} \eta_{N'} (\phi_{\mathbf{r}N' }^{\dagger} + \phi_{\mathbf{r}N' })  .
\end{split}
\end{equation}

\subsection{Bogoliubov transformation and effective interactions}
\begin{figure}[!tb]
\begin{center}
   \includegraphics[width = 0.7\columnwidth ]{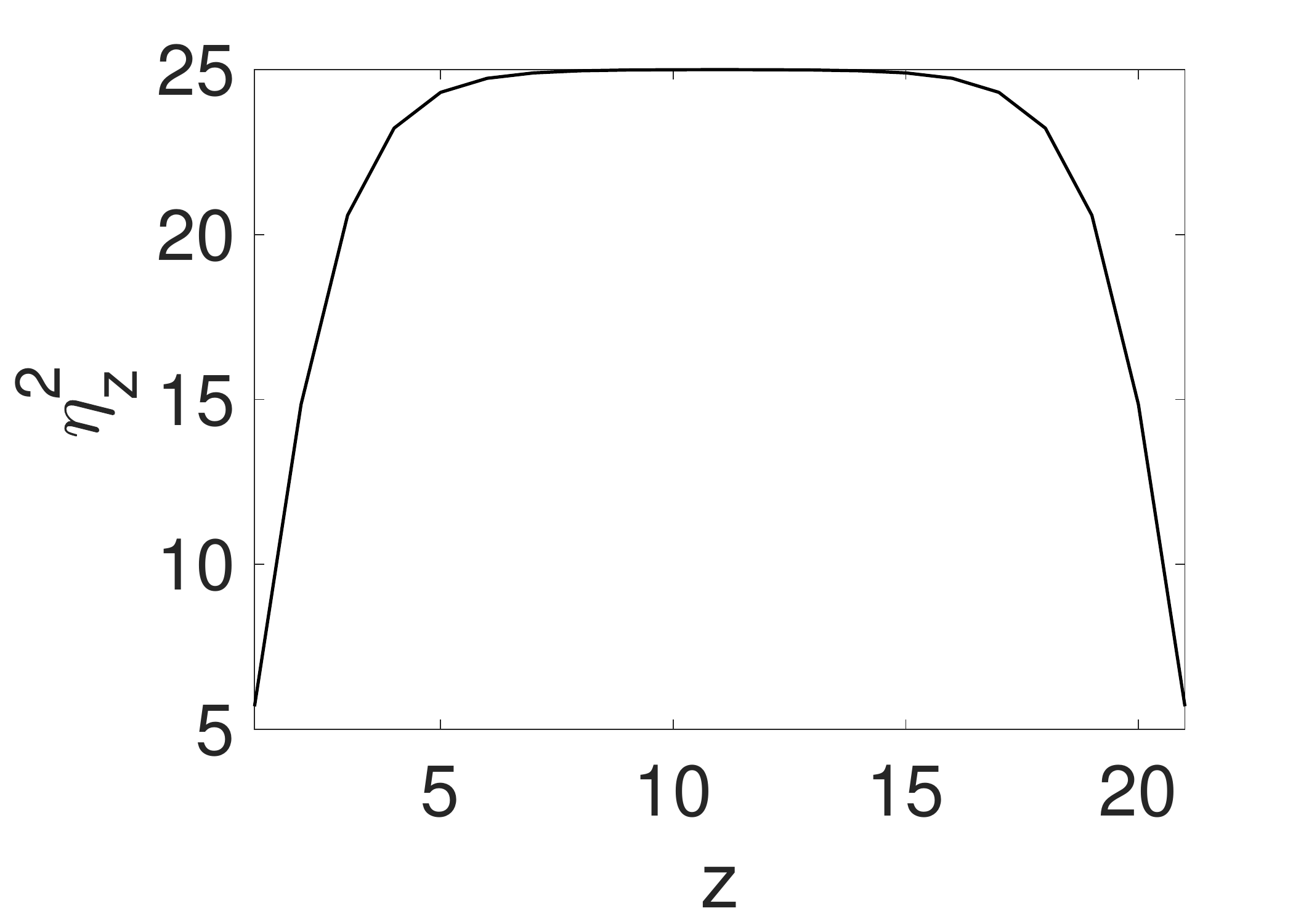}
   \caption{The spatial profile of the condensation obtained for the parameters we used for the calculations.}
\label{Nz}
\end{center}
\end{figure}

Integrating out the BEC fluctuations can be easily done in the Bogoliubov modes. For this purpose, we show the details of the Bogoliubov theory with open boundary conditions. We start from Eqs.~\eqref{HB_eff} and \eqref{GPE}. First, we solve the Gross-Pitaevskii equation numerically. Figure~\ref{Nz} shows a condensation density along the $z$ axis for $N'=11$, $t_\text{B} = 10 t$, $\mu_B = -5.5 t_\text{B}$, and $U_\text{B} = 0.2 t$. We see that the condensed density is depleted near the open boundaries, while in the center of the system the density is close to the value of a homogeneous system $\sim n_\text{B} = (6 t_\text{B} + \mu_\text{B}) /U_\text{B}$.

With the obtained density profile, we diagonalize the quadratic Hamiltonian of the fluctuating part. The Hamiltonian can be written as
\begin{equation}
H_\text{3D}^{\text{B}} = \sum_{\mathbf{p}}  \frac{1}{2} \begin{bmatrix}
\vec{\phi}_{\mathbf{p}} \\
 \vec{\phi}_{-\mathbf{p}}^{\dagger} 
\end{bmatrix}^{\dagger} 
\begin{bmatrix}
\mathbf{A}^{(\mathbf{p})} &  \mathbf{B} \\
\mathbf{B} & \mathbf{A}^{(\mathbf{p})}
\end{bmatrix}
\begin{bmatrix}
\vec{\phi}_{\mathbf{p}} \\
 \vec{\phi}_{-\mathbf{p}}^{\dagger} 
\end{bmatrix} + \text{const.},
\end{equation}
with a vector $\vec{\phi}_{\mathbf{p}} = [\phi_{\mathbf{p}1}, \dots, \phi_{\mathbf{p} ,{2N'-1}}]^T$ and $\phi_{\mathbf{p}z}$ is a partially Fourier-transformed operator along the in-plane directions. $\mathbf{A}^{(\mathbf{p})}$ is a $(2N'-1) \times (2N'-1)$ matrix that corresponds to the particle conserving part of the Hamiltonian,
\begin{multline}
\mathbf{A}^{(\mathbf{p})}_{zz'} = -t_\text{B} \left[ \delta_{z,z'-1} + \delta_{z,z'+1} \right] \\
+ \left\{ -2 t_\text{B} \left[ \cos(p_x a) +\cos(p_y a) \right] -\mu_\text{B}  + 2U_\text{B} \eta_z^2 \right\} \delta_{zz'} ,
\end {multline}
and $\mathbf{B}$ is a $(2N'-1) \times (2N'-1)$ matrix that corresponds to the particle nonconserving part of the Hamiltonian, 
\begin{equation}
\mathbf{B}_{zz'} = U_\text{B} \eta_z^2 \delta_{zz'}.
\end{equation}

The diagonalization is done by a general Bogoliubov transformation $\mathbf{W}^{(\mathbf{p})}$ for each $\mathbf{p}$ \cite{blaizot1986quantum},
\begin{equation}
\begin{bmatrix}
\vec{\phi}_{\mathbf{p}} \\
 \vec{\phi}_{-\mathbf{p}}^{\dagger} 
\end{bmatrix}
= \mathbf{W}^{(\mathbf{p})} 
\begin{bmatrix}
\vec{b}_{\mathbf{p}} \\
 \vec{b}_{-\mathbf{p}}^{\dagger} 
\end{bmatrix},
\end{equation}
leading to the final form
\begin{equation}
H_\text{3D}^{\text{B}} =  \sum_{\mathbf{p}} \sum_{\lambda=1}^{2N'-1} \omega_{\mathbf{p} \lambda} b_{\mathbf{p}\lambda}^{\dagger} b_{\mathbf{p}\lambda}
\end{equation}
with positive eigenvalues $\omega_{\mathbf{p} \lambda}$. 

\begin{figure}[!tb]
\begin{center}
   \includegraphics[width = 0.95\columnwidth ]{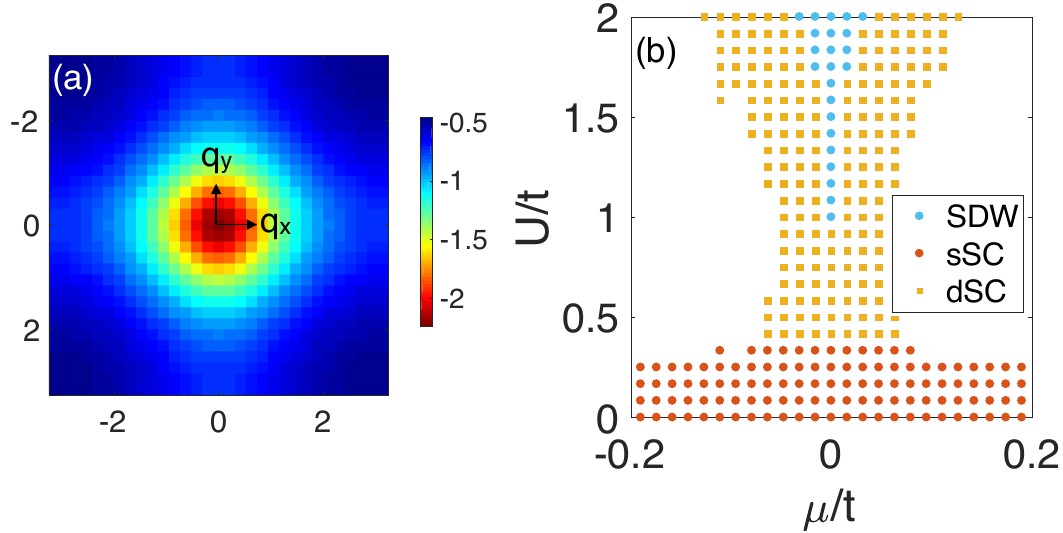}
   \caption{(a) The effective interaction $U^\text{B}_\text{eff} (\mathbf{q})$ in Eq.~\eqref{Ueff_BEC} in units of $V^2$. We take $N'=11$, $t_\text{B} = 10 t$, $\mu_B = -5.5 t_\text{B}$, and $U_\text{B} = 0.2 t$. (b) A phase diagram obtained by fRG with the same parameters as (a) and $V=0.8 t$. Blank regions correspond to Fermi liquid; no instability is found.}
\label{PD_BEC}
\end{center}
\end{figure}

Integrating out the bosons \cite{Mathey2006, Mathey2007}, we obtain the effective interaction as in Eq.~\eqref{HF_eff},
\begin{equation}
U^\text{B}_\text{eff} (\mathbf{q}) = -2 V^2 \eta_{N'}^2  \sum_{\lambda = 1}^{2N'-1}  \frac{\left( \mathbf{W}^{(\mathbf{q})}_{N' \lambda}+   \mathbf{W}^{(\mathbf{q})}_{3N'-1, \lambda} \right)^2 }{\omega_{\mathbf{q} \lambda}}.
\label{Ueff_BEC}
\end{equation}
An example of the effective interaction $U^\text{B}_\text{eff} (\mathbf{q})$ is plotted in Fig.~\ref{PD_BEC} (a). The strong negative peak at $\mathbf{q} = 0$ indicates attractive on-site and NN interactions.

For the spatially homogeneous case or when the boundary effects can be ignored, the real space interaction decays exponentially, and the interaction range is roughly the healing length of the BEC, $\xi = \sqrt{t_\text{B}/2n_\text{B} U_\text{B}} \sim 2.2 a$ \cite{Illuminati2004, Mathey2006}. In our setup, we find that the induced interaction $U^\text{B}_\text{eff} (\mathbf{r})$ also decays quickly in real space over a few lattice constants. Therefore, the essential features of the induced interaction do not depend on the specifics of the system as long as we assume a well-defined BEC.

\subsection{Phase diagrams}
The zero-temperature phase diagram obtained by fRG is given in Fig.~\ref{PD_BEC}(b). We take $N'=11$, $t_\text{B} = 10 t$, $\mu_B = -5.5 t_\text{B}$, $U_\text{B} = 0.2 t $, and $V=0.8 t$. The induced interaction reduces the on-site interaction as
\begin{equation}
U \rightarrow U -|U^\text{B}_\text{eff} (\mathbf{r}=0)| ,
\end{equation}
and thus for $U < |U^\text{B}_\text{eff} (\mathbf{r}=0)| \approx 0.33t$, the model is reduced to the attractive Hubbard model leading to $s$-wave SC. When $U > |U^\text{B}_\text{eff} (\mathbf{r}=0)|$, the model is nearly identical to the repulsive Hubbard model except for negligible attractive NN interactions. Thus, at half-filling ($\mu=0$) we find SDW for large $U$, and $d_{x^2-y^2}$-wave SC for intermediate $U$. Slightly away from half-filling, the system shows $d_{x^2-y^2}$-wave pairing, which disappears as the filling deviates further away from half-filling. This phase diagram is qualitatively similar to the one in Ref.~\onlinecite{Mathey2006} obtained for a 2D Bose-Fermi mixture.

\section{Discussions}
\label{discussion}
\subsection{Critical temperatures}
\begin{figure}[tb]
\begin{center}
   \includegraphics[width = 0.95\columnwidth ]{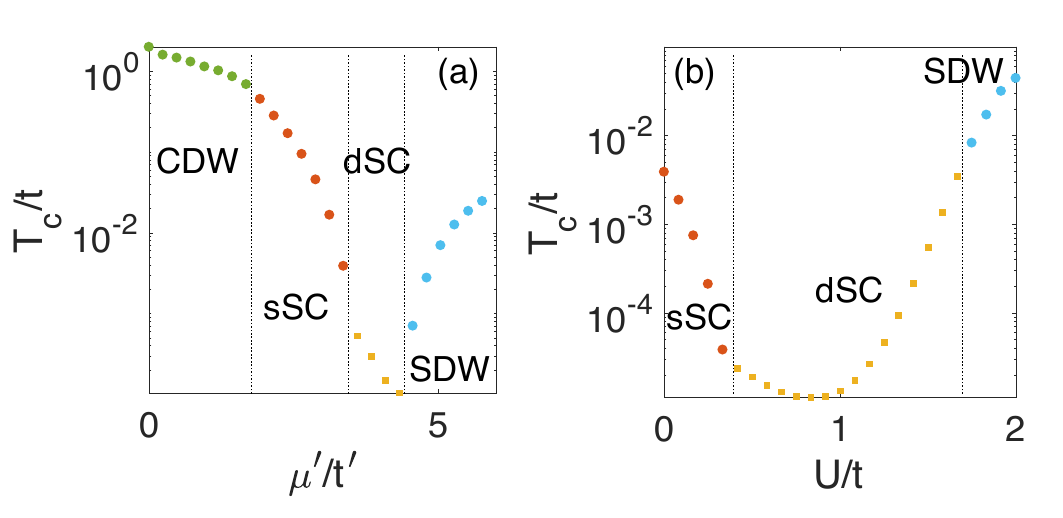}
   \caption{(a) The critical temperatures of phases found in the Fermi-Fermi mixture at half-filling and $V'/t'=1$ as in Fig.~\ref{PD_Fermi}(a). (b) The critical temperatures of phases found in the Fermi-BEC mixture at $\mu = -0.016$ as in Fig.~\ref{PD_BEC}(b). The vertical dotted lines are phase boundaries.}
\label{Tc}
\end{center}
\end{figure}

Here we discuss the critical temperatures $T_c$ or gap energies of ordering phases, which are estimated from the ultraviolet cutoff energy where these instabilities occur (We note that the determination of $T_c$ depends on minor complications such as when we stop the RG flows \cite{Reiss2007, Yamase2016}.) Figure \ref{Tc}(a) shows the critical temperatures for the Fermi-Fermi mixture at $V/t'=1$ and $U = 1$ at half-filling [see Fig.~\ref{PD_Fermi}(a)] for different 3D chemical potentials. The critical temperatures of SDW, CDW, and $s$-wave SC are comparable to the Fermi energy $\sim t$, and it seems possible to detect these orders experimentally \cite{Parsons2016}.

On the other hand, the critical temperature of $d$-wave SC is rather small, $T_c \sim 5.0\times10^{-3} t$. However, we note that this value is an order of magnitude bigger than that of the same pairing induced by doping in the simple Hubbard model with only on-site interaction $U$. For example, at $\mu = -0.025$ with $U=1$ and $V=0$ [see Fig.~\ref{PD_Fermi}(b)], we find the gap energy of the $d$-wave SC as $5.0\times 10^{-5} t$. While we expect that the gap of the interaction-induced $d$-wave SC depends on the NN interaction strength, our finding indicates that the $d$-wave ordering induced by the NN interaction is more stable than the one induced by doping within the regime we have studied. In the simple Hubbard model, it is known that the NNN hopping enhances the $d$-wave pairing by destroying the perfect nesting at $\mathbf{q} = (\pi,\pi)$ \cite{Halboth2000, Honerkamp2001, Honerkamp2001a}. Similarly in our model, we expect that the NNN hopping may further enhance the gap energy of the $d$-wave SC induced by the NN interaction.

Figure~\ref{Tc}(b) shows the critical temperatures for the Fermi-BEC mixture at slightly away from half-filling $\mu = -0.016$ and $V=0.8t$ with various $U$ [see Fig.~\ref{PD_BEC}(b)]. The maximal critical temperature of the $d$-wave SC, $T_c \sim 5.0\times 10^{-3} t$, is comparable to that of the Fermi-Fermi mixture at half-filling.

\subsection{Experimental perspectives}
One of the important control parameters in our model is the interspecies interaction $V$. We propose that the following atomic mixtures in species-specific optical lattices can be employed to realize our setup with control over $V$. For Fermi-Fermi mixtures, recent experiments have achieved control of the interaction strength from the weak to strong interactions by interspecies Feshbach resonances for $^6$Li-$^{40}$K mixtures \cite{Wille2008, Voigt2009,Spiegelhalder2009, Tiecke2010,Spiegelhalder2010}. For Fermi-Bose scattering, a $^{23}$Na-$^{40}$K mixture \cite{Park2012} and a $^{87}$Rb-$^{40}$K mixture \cite{Ospelkaus2006, Ospelkaus2006a, Best2009} show Feshbach resonances. In other combinations, the $s$-wave triplet scattering length $\sim$ $20 a_B$ between $^{87}$Rb and $^{87}$Li has been measured \cite{Silber2005}. A scattering length $ \sim 13 a_B$ was observed in a $^6$Li-$^{174}$Yi mixture \cite{Ivanov2011,Hara2011}. Isotopes of Yb are also used to realize a Bose-Fermi-Hubbard system with $V \sim 100 t$ in Ref.~\onlinecite{Sugawa2010}. 

Finally, we note that the various high angular-momentum pairings that we find may be detected experimentally by, for example, phase-sensitive measurements \cite{Carusotto2005, Dao2007, Gritsev2008, Pekker2009}, or using noise correlations \cite{Altman2004, Kitagawa2011}.

\section{Conclusions}
\label{conclusion}
In this paper, we have investigated the mediated pairing in a 2D Fermi gas embedded in a 3D system, either a noninteracting Fermi gas or a BEC. The induced interaction among 2D fermions obtained by integrating out the 3D degrees of freedom lead to various pairing instabilities such as $s$-wave, $p$-wave, $d_{x^2-y^2}$-wave, and $g_{xy(x^2- y^2)}$-wave superconductivity. In particular, we have shown that by using the 3D fermions, we can explore various parameter regimes of the extended Hubbard model, once we map the induced interaction onto on-site and nearest-neighbor interactions. 

We also find that the $d$-wave superconductivity induced by the attractive nearest-neighbor interaction can have a higher critical temperature than that of the same pairing induced by doping in the simple Hubbard model. The former mechanism, combined with the next-nearest hopping (known to enhance the $d$-wave pairing instability), may be used to realize this exotic pairing in cold atom experiments.

\acknowledgements
J.O. and L.M. acknowledge the support from the Deutsche Forschungsgemeinschaft (through SFB 925 and EXC 1074) and from the Landesexzellenzinitiative Hamburg, which is supported by the Joachim Herz Stiftung. W.M.H. acknowledges support from the Ministry of Science and Technology in Taiwan through Grant No. MOST 104-2112-M-005-006-MY3.

\appendix
\section{two-dimensional limit: effective interactions and phase diagrams}
\begin{figure}[!tb]
\begin{center}
   \includegraphics[width = 0.9\columnwidth ]{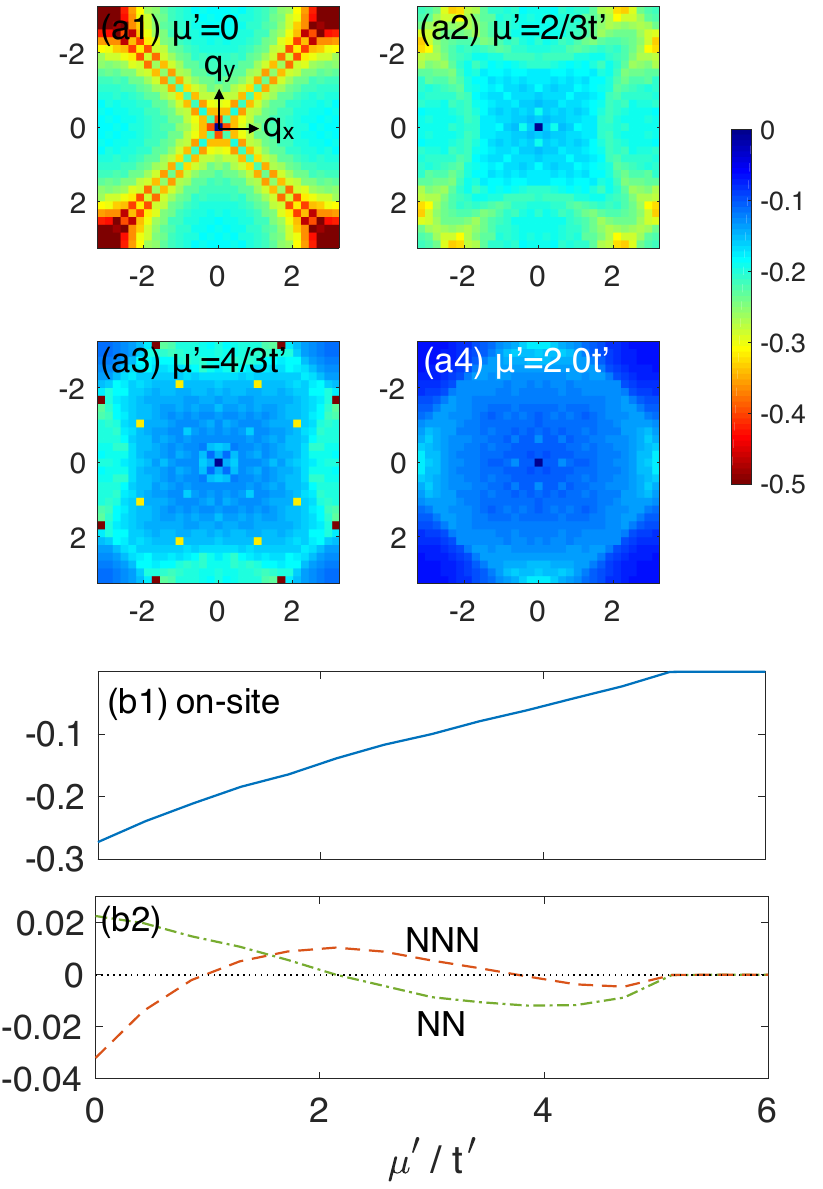}
   \caption{(a) The effective interaction in momentum space in units of $4V^2/t'$. We use $N=50$ and $N' = 1$. (a1)-(a4) represent the dependence of $U^\text{F}_\text{eff} (\mathbf{q})$ on $\mu'$. (b) The effective interactions in real space in units of $4V^2/t'$. }
\label{Ueff_SM}
\end{center}
\end{figure}

\begin{figure}[!tb]
\begin{center}
   \includegraphics[width = 0.9\columnwidth ]{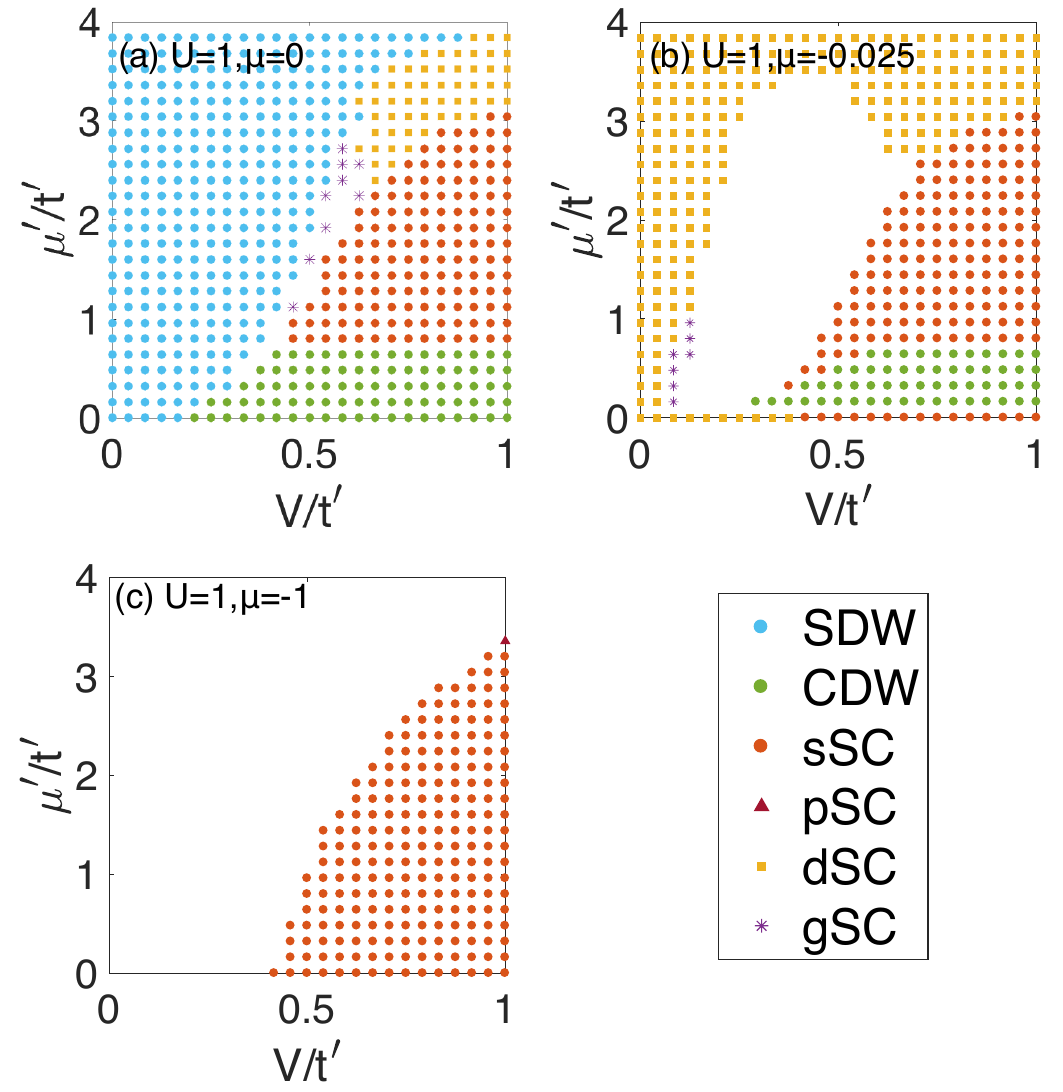}
   \caption{Phase diagrams of a 2D Fermi gas in contact with a noninteracting 2D Fermi gas. We use $N=50$, $N'=1$, and $t' = 8t$. Blank regions correspond to Fermi liquid; no instability is found.}
\label{PD_SM}
\end{center}
\end{figure}

In this Appendix, we discuss the case in which the bulk system is in the two-dimensional limit, $N'=1$. The effective interaction in momentum space and in real space is depicted in Fig.~\ref{Ueff_SM}. We find that the scattering along the diagonal directions $|q_x| = |q_y|$ are enhanced near half-filling compared to the bulk case ($N'=11$) where we only have a sharp peak at $\mathbf{q} = (\pi, \pi)$. This results in larger next-nearest-neighbor interactions as shown in Fig.~\ref{Ueff_SM}(b). However, the overall tendency and dependence on the filling is quite similar to the bulk case, and the obtained phase diagrams shown in Fig.~\ref{PD_SM} are also qualitatively the same as the bulk case. This is because that the Lindhard function or particle-hole propagator in Eq.~\eqref{UE_eff} shows oscillatory behaviors regardless of the dimensionality. While for the 3D case we need to project it onto the 2D plane to get the effective interaction, this does not modify the oscillatory nature of the function qualitatively.

We note that when the bulk system is truly two-dimensional, the effect of $V$ on the bulk part and of retardation effect may not be ignorable due to larger quantum fluctuations, and our approximations need to be carefully examined. Treating both atomic species by fRG as in Ref.~\onlinecite{Lai2014} seems more appropriate in this case.


\begin{thebibliography}{73}
\expandafter\ifx\csname natexlab\endcsname\relax\def\natexlab#1{#1}\fi
\expandafter\ifx\csname bibnamefont\endcsname\relax
  \def\bibnamefont#1{#1}\fi
\expandafter\ifx\csname bibfnamefont\endcsname\relax
  \def\bibfnamefont#1{#1}\fi
\expandafter\ifx\csname citenamefont\endcsname\relax
  \def\citenamefont#1{#1}\fi
\expandafter\ifx\csname url\endcsname\relax
  \def\url#1{\texttt{#1}}\fi
\expandafter\ifx\csname urlprefix\endcsname\relax\def\urlprefix{URL }\fi
\providecommand{\bibinfo}[2]{#2}
\providecommand{\eprint}[2][]{\url{#2}}

\bibitem[{\citenamefont{Lewenstein et~al.}(2007)\citenamefont{Lewenstein,
  Sanpera, Ahufinger, Damski, Sen(De), and Sen}}]{Lewenstein2007}
\bibinfo{author}{\bibfnamefont{M.}~\bibnamefont{Lewenstein}},
  \bibinfo{author}{\bibfnamefont{A.}~\bibnamefont{Sanpera}},
  \bibinfo{author}{\bibfnamefont{V.}~\bibnamefont{Ahufinger}},
  \bibinfo{author}{\bibfnamefont{B.}~\bibnamefont{Damski}},
  \bibinfo{author}{\bibfnamefont{A.}~\bibnamefont{Sen(De)}}, \bibnamefont{and}
  \bibinfo{author}{\bibfnamefont{U.}~\bibnamefont{Sen}}, \bibinfo{journal}{Adv.
  Phys.} \textbf{\bibinfo{volume}{56}}, \bibinfo{pages}{243}
  (\bibinfo{year}{2007}).

\bibitem[{\citenamefont{Bloch et~al.}(2008)\citenamefont{Bloch, Dalibard, and
  Zwerger}}]{Bloch2008}
\bibinfo{author}{\bibfnamefont{I.}~\bibnamefont{Bloch}},
  \bibinfo{author}{\bibfnamefont{J.}~\bibnamefont{Dalibard}}, \bibnamefont{and}
  \bibinfo{author}{\bibfnamefont{W.}~\bibnamefont{Zwerger}},
  \bibinfo{journal}{Rev. Mod. Phys.} \textbf{\bibinfo{volume}{80}},
  \bibinfo{pages}{885} (\bibinfo{year}{2008}).

\bibitem[{\citenamefont{K{\"{o}}hl et~al.}(2005)\citenamefont{K{\"{o}}hl,
  Moritz, St{\"{o}}ferle, G{\"{u}}nter, and Esslinger}}]{Kohl2005}
\bibinfo{author}{\bibfnamefont{M.}~\bibnamefont{K{\"{o}}hl}},
  \bibinfo{author}{\bibfnamefont{H.}~\bibnamefont{Moritz}},
  \bibinfo{author}{\bibfnamefont{T.}~\bibnamefont{St{\"{o}}ferle}},
  \bibinfo{author}{\bibfnamefont{K.}~\bibnamefont{G{\"{u}}nter}},
  \bibnamefont{and}
  \bibinfo{author}{\bibfnamefont{T.}~\bibnamefont{Esslinger}},
  \bibinfo{journal}{Phys. Rev. Lett.} \textbf{\bibinfo{volume}{94}},
  \bibinfo{pages}{080403} (\bibinfo{year}{2005}).

\bibitem[{\citenamefont{Esslinger}(2010)}]{Esslinger2010}
\bibinfo{author}{\bibfnamefont{T.}~\bibnamefont{Esslinger}},
  \bibinfo{journal}{Annu. Rev. Condens. Matter Phys.}
  \textbf{\bibinfo{volume}{1}}, \bibinfo{pages}{129} (\bibinfo{year}{2010}).

\bibitem[{\citenamefont{Jaksch et~al.}(1998)\citenamefont{Jaksch, Bruder,
  Cirac, Gardiner, and Zoller}}]{Jaksch1998}
\bibinfo{author}{\bibfnamefont{D.}~\bibnamefont{Jaksch}},
  \bibinfo{author}{\bibfnamefont{C.}~\bibnamefont{Bruder}},
  \bibinfo{author}{\bibfnamefont{J.~I.} \bibnamefont{Cirac}},
  \bibinfo{author}{\bibfnamefont{C.~W.} \bibnamefont{Gardiner}},
  \bibnamefont{and} \bibinfo{author}{\bibfnamefont{P.}~\bibnamefont{Zoller}},
  \bibinfo{journal}{Phys. Rev. Lett.} \textbf{\bibinfo{volume}{81}},
  \bibinfo{pages}{3108} (\bibinfo{year}{1998}).

\bibitem[{\citenamefont{Greiner et~al.}(2002)\citenamefont{Greiner, Mandel,
  Esslinger, H{\"{a}}nsch, and Bloch}}]{Greiner2002}
\bibinfo{author}{\bibfnamefont{M.}~\bibnamefont{Greiner}},
  \bibinfo{author}{\bibfnamefont{O.}~\bibnamefont{Mandel}},
  \bibinfo{author}{\bibfnamefont{T.}~\bibnamefont{Esslinger}},
  \bibinfo{author}{\bibfnamefont{T.~W.} \bibnamefont{H{\"{a}}nsch}},
  \bibnamefont{and} \bibinfo{author}{\bibfnamefont{I.}~\bibnamefont{Bloch}},
  \bibinfo{journal}{Nature (London)} \textbf{\bibinfo{volume}{415}}, \bibinfo{pages}{39}
  (\bibinfo{year}{2002}).

\bibitem[{\citenamefont{Jaksch and Zoller}(2005)}]{Jaksch2005}
\bibinfo{author}{\bibfnamefont{D.}~\bibnamefont{Jaksch}} \bibnamefont{and}
  \bibinfo{author}{\bibfnamefont{P.}~\bibnamefont{Zoller}},
  \bibinfo{journal}{Ann. Phys. (N. Y).} \textbf{\bibinfo{volume}{315}},
  \bibinfo{pages}{52} (\bibinfo{year}{2005}).

\bibitem[{\citenamefont{Lahaye et~al.}(2009)\citenamefont{Lahaye, Menotti,
  Santos, Lewenstein, and Pfau}}]{Lahaye2009}
\bibinfo{author}{\bibfnamefont{T.}~\bibnamefont{Lahaye}},
  \bibinfo{author}{\bibfnamefont{C.}~\bibnamefont{Menotti}},
  \bibinfo{author}{\bibfnamefont{L.}~\bibnamefont{Santos}},
  \bibinfo{author}{\bibfnamefont{M.}~\bibnamefont{Lewenstein}},
  \bibnamefont{and} \bibinfo{author}{\bibfnamefont{T.}~\bibnamefont{Pfau}},
  \bibinfo{journal}{Reports Prog. Phys.} \textbf{\bibinfo{volume}{72}},
  \bibinfo{pages}{126401} (\bibinfo{year}{2009}).

\bibitem[{\citenamefont{Dutta et~al.}(2015)\citenamefont{Dutta, Gajda, Hauke,
  Lewenstein, L{\"{u}}hmann, Malomed, Sowi{\'{n}}ski, and
  Zakrzewski}}]{Dutta2015}
\bibinfo{author}{\bibfnamefont{O.}~\bibnamefont{Dutta}},
  \bibinfo{author}{\bibfnamefont{M.}~\bibnamefont{Gajda}},
  \bibinfo{author}{\bibfnamefont{P.}~\bibnamefont{Hauke}},
  \bibinfo{author}{\bibfnamefont{M.}~\bibnamefont{Lewenstein}},
  \bibinfo{author}{\bibfnamefont{D.-S.} \bibnamefont{L{\"{u}}hmann}},
  \bibinfo{author}{\bibfnamefont{B.~A.} \bibnamefont{Malomed}},
  \bibinfo{author}{\bibfnamefont{T.}~\bibnamefont{Sowi{\'{n}}ski}},
  \bibnamefont{and}
  \bibinfo{author}{\bibfnamefont{J.}~\bibnamefont{Zakrzewski}},
  \bibinfo{journal}{Reports Prog. Phys.} \textbf{\bibinfo{volume}{78}},
  \bibinfo{pages}{066001} (\bibinfo{year}{2015}).

\bibitem[{\citenamefont{Johnson and Rolston}(2010)}]{Johnson2010}
\bibinfo{author}{\bibfnamefont{J.~E.} \bibnamefont{Johnson}} \bibnamefont{and}
  \bibinfo{author}{\bibfnamefont{S.~L.} \bibnamefont{Rolston}},
  \bibinfo{journal}{Phys. Rev. A} \textbf{\bibinfo{volume}{82}},
  \bibinfo{pages}{033412} (\bibinfo{year}{2010}).

\bibitem[{\citenamefont{Pupillo et~al.}(2010)\citenamefont{Pupillo, Micheli,
  Boninsegni, Lesanovsky, and Zoller}}]{Pupillo2010}
\bibinfo{author}{\bibfnamefont{G.}~\bibnamefont{Pupillo}},
  \bibinfo{author}{\bibfnamefont{A.}~\bibnamefont{Micheli}},
  \bibinfo{author}{\bibfnamefont{M.}~\bibnamefont{Boninsegni}},
  \bibinfo{author}{\bibfnamefont{I.}~\bibnamefont{Lesanovsky}},
  \bibnamefont{and} \bibinfo{author}{\bibfnamefont{P.}~\bibnamefont{Zoller}},
  \bibinfo{journal}{Phys. Rev. Lett.} \textbf{\bibinfo{volume}{104}},
  \bibinfo{pages}{223002} (\bibinfo{year}{2010}).

\bibitem[{\citenamefont{Henkel et~al.}(2010)\citenamefont{Henkel, Nath, and
  Pohl}}]{Henkel2010}
\bibinfo{author}{\bibfnamefont{N.}~\bibnamefont{Henkel}},
  \bibinfo{author}{\bibfnamefont{R.}~\bibnamefont{Nath}}, \bibnamefont{and}
  \bibinfo{author}{\bibfnamefont{T.}~\bibnamefont{Pohl}},
  \bibinfo{journal}{Phys. Rev. Lett.} \textbf{\bibinfo{volume}{104}},
  \bibinfo{pages}{195302} (\bibinfo{year}{2010}).

\bibitem[{\citenamefont{Honer et~al.}(2010)\citenamefont{Honer, Weimer, Pfau,
  and B{\"{u}}chler}}]{Honer2010}
\bibinfo{author}{\bibfnamefont{J.}~\bibnamefont{Honer}},
  \bibinfo{author}{\bibfnamefont{H.}~\bibnamefont{Weimer}},
  \bibinfo{author}{\bibfnamefont{T.}~\bibnamefont{Pfau}}, \bibnamefont{and}
  \bibinfo{author}{\bibfnamefont{H.~P.} \bibnamefont{B{\"{u}}chler}},
  \bibinfo{journal}{Phys. Rev. Lett.} \textbf{\bibinfo{volume}{105}},
  \bibinfo{pages}{160404} (\bibinfo{year}{2010}).

\bibitem[{\citenamefont{Balewski et~al.}(2014)\citenamefont{Balewski, Krupp,
  Gaj, Hofferberth, L{\"{o}}w, and Pfau}}]{Balewski2014}
\bibinfo{author}{\bibfnamefont{J.~B.} \bibnamefont{Balewski}},
  \bibinfo{author}{\bibfnamefont{A.~T.} \bibnamefont{Krupp}},
  \bibinfo{author}{\bibfnamefont{A.}~\bibnamefont{Gaj}},
  \bibinfo{author}{\bibfnamefont{S.}~\bibnamefont{Hofferberth}},
  \bibinfo{author}{\bibfnamefont{R.}~\bibnamefont{L{\"{o}}w}},
  \bibnamefont{and} \bibinfo{author}{\bibfnamefont{T.}~\bibnamefont{Pfau}},
  \bibinfo{journal}{New J. Phys.} \textbf{\bibinfo{volume}{16}},
  \bibinfo{pages}{063012} (\bibinfo{year}{2014}).

\bibitem[{\citenamefont{Scalapino}(2012)}]{Scalapino2012}
\bibinfo{author}{\bibfnamefont{D.~J.} \bibnamefont{Scalapino}},
  \bibinfo{journal}{Rev. Mod. Phys.} \textbf{\bibinfo{volume}{84}},
  \bibinfo{pages}{1383} (\bibinfo{year}{2012}).

\bibitem[{\citenamefont{Huang et~al.}(2013{\natexlab{a}})\citenamefont{Huang,
  Lai, Shi, and Tsai}}]{Huang2013a}
\bibinfo{author}{\bibfnamefont{W.-M.} \bibnamefont{Huang}},
  \bibinfo{author}{\bibfnamefont{C.~Y.} \bibnamefont{Lai}},
  \bibinfo{author}{\bibfnamefont{C.}~\bibnamefont{Shi}}, \bibnamefont{and}
  \bibinfo{author}{\bibfnamefont{S.-W.} \bibnamefont{Tsai}},
  \bibinfo{journal}{Phys. Rev. B} \textbf{\bibinfo{volume}{88}},
  \bibinfo{pages}{054504} (\bibinfo{year}{2013}{\natexlab{a}}).

\bibitem[{\citenamefont{LeBlanc and Thywissen}(2007)}]{LeBlanc2007}
\bibinfo{author}{\bibfnamefont{L.~J.} \bibnamefont{LeBlanc}} \bibnamefont{and}
  \bibinfo{author}{\bibfnamefont{J.~H.} \bibnamefont{Thywissen}},
  \bibinfo{journal}{Phys. Rev. A} \textbf{\bibinfo{volume}{75}},
  \bibinfo{pages}{053612} (\bibinfo{year}{2007}).

\bibitem[{\citenamefont{Catani et~al.}(2009)\citenamefont{Catani, Barontini,
  Lamporesi, Rabatti, Thalhammer, Minardi, Stringari, and
  Inguscio}}]{Catani2009}
\bibinfo{author}{\bibfnamefont{J.}~\bibnamefont{Catani}},
  \bibinfo{author}{\bibfnamefont{G.}~\bibnamefont{Barontini}},
  \bibinfo{author}{\bibfnamefont{G.}~\bibnamefont{Lamporesi}},
  \bibinfo{author}{\bibfnamefont{F.}~\bibnamefont{Rabatti}},
  \bibinfo{author}{\bibfnamefont{G.}~\bibnamefont{Thalhammer}},
  \bibinfo{author}{\bibfnamefont{F.}~\bibnamefont{Minardi}},
  \bibinfo{author}{\bibfnamefont{S.}~\bibnamefont{Stringari}},
  \bibnamefont{and} \bibinfo{author}{\bibfnamefont{M.}~\bibnamefont{Inguscio}},
  \bibinfo{journal}{Phys. Rev. Lett.} \textbf{\bibinfo{volume}{103}},
  \bibinfo{pages}{140401} (\bibinfo{year}{2009}).

\bibitem[{\citenamefont{Lamporesi et~al.}(2010)\citenamefont{Lamporesi, Catani,
  Barontini, Nishida, Inguscio, and Minardi}}]{Lamporesi2010}
\bibinfo{author}{\bibfnamefont{G.}~\bibnamefont{Lamporesi}},
  \bibinfo{author}{\bibfnamefont{J.}~\bibnamefont{Catani}},
  \bibinfo{author}{\bibfnamefont{G.}~\bibnamefont{Barontini}},
  \bibinfo{author}{\bibfnamefont{Y.}~\bibnamefont{Nishida}},
  \bibinfo{author}{\bibfnamefont{M.}~\bibnamefont{Inguscio}}, \bibnamefont{and}
  \bibinfo{author}{\bibfnamefont{F.}~\bibnamefont{Minardi}},
  \bibinfo{journal}{Phys. Rev. Lett.} \textbf{\bibinfo{volume}{104}},
  \bibinfo{pages}{153202} (\bibinfo{year}{2010}).

\bibitem[{\citenamefont{Haller et~al.}(2010)\citenamefont{Haller, Mark, Hart,
  Danzl, Reichs{\"{o}}llner, Melezhik, Schmelcher, and
  N{\"{a}}gerl}}]{Haller2010}
\bibinfo{author}{\bibfnamefont{E.}~\bibnamefont{Haller}},
  \bibinfo{author}{\bibfnamefont{M.~J.} \bibnamefont{Mark}},
  \bibinfo{author}{\bibfnamefont{R.}~\bibnamefont{Hart}},
  \bibinfo{author}{\bibfnamefont{J.~G.} \bibnamefont{Danzl}},
  \bibinfo{author}{\bibfnamefont{L.}~\bibnamefont{Reichs{\"{o}}llner}},
  \bibinfo{author}{\bibfnamefont{V.}~\bibnamefont{Melezhik}},
  \bibinfo{author}{\bibfnamefont{P.}~\bibnamefont{Schmelcher}},
  \bibnamefont{and} \bibinfo{author}{\bibfnamefont{H.-C.}
  \bibnamefont{N{\"{a}}gerl}}, \bibinfo{journal}{Physcal Rev. Lett.}
  \textbf{\bibinfo{volume}{104}}, \bibinfo{pages}{153203}
  (\bibinfo{year}{2010}).

\bibitem[{\citenamefont{Nishida and Tan}(2008)}]{Nishida2008}
\bibinfo{author}{\bibfnamefont{Y.}~\bibnamefont{Nishida}} \bibnamefont{and}
  \bibinfo{author}{\bibfnamefont{S.}~\bibnamefont{Tan}},
  \bibinfo{journal}{Phys. Rev. Lett.} \textbf{\bibinfo{volume}{101}},
  \bibinfo{pages}{170401} (\bibinfo{year}{2008}).

\bibitem[{\citenamefont{Nishida and Tan}(2010)}]{Nishida2010a}
\bibinfo{author}{\bibfnamefont{Y.}~\bibnamefont{Nishida}} \bibnamefont{and}
  \bibinfo{author}{\bibfnamefont{S.}~\bibnamefont{Tan}},
  \bibinfo{journal}{Phys. Rev. A} \textbf{\bibinfo{volume}{82}},
  \bibinfo{pages}{062713} (\bibinfo{year}{2010}).

\bibitem[{\citenamefont{Iskin and Subasi}(2010)}]{Iskin2010}
\bibinfo{author}{\bibfnamefont{M.}~\bibnamefont{Iskin}} \bibnamefont{and}
  \bibinfo{author}{\bibfnamefont{A.~L.} \bibnamefont{Subasi}},
  \bibinfo{journal}{Phys. Rev. A} \textbf{\bibinfo{volume}{82}},
  \bibinfo{pages}{063628} (\bibinfo{year}{2010}).

\bibitem[{\citenamefont{Yang et~al.}(2011)\citenamefont{Yang, Huang, and
  Wan}}]{Yang2011}
\bibinfo{author}{\bibfnamefont{X.~S.} \bibnamefont{Yang}},
  \bibinfo{author}{\bibfnamefont{B.~B.} \bibnamefont{Huang}}, \bibnamefont{and}
  \bibinfo{author}{\bibfnamefont{S.~L.} \bibnamefont{Wan}},
  \bibinfo{journal}{Eur. Phys. J. B} \textbf{\bibinfo{volume}{83}},
  \bibinfo{pages}{445} (\bibinfo{year}{2011}).

\bibitem[{\citenamefont{Huang et~al.}(2013{\natexlab{b}})\citenamefont{Huang,
  Irwin, and Tsai}}]{Huang2013}
\bibinfo{author}{\bibfnamefont{W.~M.} \bibnamefont{Huang}},
  \bibinfo{author}{\bibfnamefont{K.}~\bibnamefont{Irwin}}, \bibnamefont{and}
  \bibinfo{author}{\bibfnamefont{S.-W.} \bibnamefont{Tsai}},
  \bibinfo{journal}{Phys. Rev. A} \textbf{\bibinfo{volume}{87}},
  \bibinfo{pages}{031603(R)} (\bibinfo{year}{2013}{\natexlab{b}}).

\bibitem[{\citenamefont{Kim et~al.}(2013)\citenamefont{Kim, Lehikoinen, and
  T{\"{o}}rm{\"{a}}}}]{Kim2013}
\bibinfo{author}{\bibfnamefont{D.~H.} \bibnamefont{Kim}},
  \bibinfo{author}{\bibfnamefont{J.~S.~J.} \bibnamefont{Lehikoinen}},
  \bibnamefont{and}
  \bibinfo{author}{\bibfnamefont{P.}~\bibnamefont{T{\"{o}}rm{\"{a}}}},
  \bibinfo{journal}{Phys. Rev. Lett.} \textbf{\bibinfo{volume}{110}},
  \bibinfo{pages}{055301} (\bibinfo{year}{2013}).

\bibitem[{\citenamefont{Young-S et~al.}(2010)\citenamefont{Young-S, Salasnich,
  and Adhikari}}]{Young2010}
\bibinfo{author}{\bibfnamefont{Luis~E.} \bibnamefont{Young-S}},
  \bibinfo{author}{\bibfnamefont{L.}~\bibnamefont{Salasnich}},
  \bibnamefont{and} \bibinfo{author}{\bibfnamefont{S.~K.}
  \bibnamefont{Adhikari}}, \bibinfo{journal}{Phys. Rev. A}
  \textbf{\bibinfo{volume}{82}}, \bibinfo{pages}{053601}
  (\bibinfo{year}{2010}).

\bibitem[{\citenamefont{Minardi et~al.}(2011)\citenamefont{Minardi, Barontini,
  Catani, Lamporesi, Nishida, and Inguscio}}]{Minardi2011}
\bibinfo{author}{\bibfnamefont{F.}~\bibnamefont{Minardi}},
  \bibinfo{author}{\bibfnamefont{G.}~\bibnamefont{Barontini}},
  \bibinfo{author}{\bibfnamefont{J.}~\bibnamefont{Catani}},
  \bibinfo{author}{\bibfnamefont{G.}~\bibnamefont{Lamporesi}},
  \bibinfo{author}{\bibfnamefont{Y.}~\bibnamefont{Nishida}}, \bibnamefont{and}
  \bibinfo{author}{\bibfnamefont{M.}~\bibnamefont{Inguscio}},
  \bibinfo{journal}{J. Phys. Conf. Ser.} \textbf{\bibinfo{volume}{264}},
  \bibinfo{pages}{012016} (\bibinfo{year}{2011}).

\bibitem[{\citenamefont{Malatsetxebarria
  et~al.}(2013{\natexlab{a}})\citenamefont{Malatsetxebarria, Cai,
  Schollw{\"{o}}ck, and Cazalilla}}]{Malatsetxebarria2013}
\bibinfo{author}{\bibfnamefont{E.}~\bibnamefont{Malatsetxebarria}},
  \bibinfo{author}{\bibfnamefont{Z.}~\bibnamefont{Cai}},
  \bibinfo{author}{\bibfnamefont{U.}~\bibnamefont{Schollw{\"{o}}ck}},
  \bibnamefont{and} \bibinfo{author}{\bibfnamefont{M.~A.}
  \bibnamefont{Cazalilla}}, \bibinfo{journal}{Phys. Rev. A}
  \textbf{\bibinfo{volume}{88}}, \bibinfo{pages}{063630}
  (\bibinfo{year}{2013}{\natexlab{a}}).

\bibitem[{\citenamefont{Malatsetxebarria
  et~al.}(2013{\natexlab{b}})\citenamefont{Malatsetxebarria, Marchetti, and
  Cazalilla}}]{Malatsetxebarria2013a}
\bibinfo{author}{\bibfnamefont{E.}~\bibnamefont{Malatsetxebarria}},
  \bibinfo{author}{\bibfnamefont{F.~M.} \bibnamefont{Marchetti}},
  \bibnamefont{and} \bibinfo{author}{\bibfnamefont{M.~A.}
  \bibnamefont{Cazalilla}}, \bibinfo{journal}{Phys. Rev. A}
  \textbf{\bibinfo{volume}{88}}, \bibinfo{pages}{033604}
  (\bibinfo{year}{2013}{\natexlab{b}}).

\bibitem[{\citenamefont{Wu and Bruun}(2016)}]{Wu2016}
\bibinfo{author}{\bibfnamefont{Z.}~\bibnamefont{Wu}} \bibnamefont{and}
  \bibinfo{author}{\bibfnamefont{G.~M.} \bibnamefont{Bruun}},
  \bibinfo{journal}{Phys. Rev. Lett.} \textbf{\bibinfo{volume}{117}},
  \bibinfo{pages}{245302} (\bibinfo{year}{2016}).

\bibitem[{\citenamefont{Midtgaard et~al.}(2016)\citenamefont{Midtgaard, Wu, and
  Bruun}}]{Midtgaard2016}
\bibinfo{author}{\bibfnamefont{J.~M.} \bibnamefont{Midtgaard}},
  \bibinfo{author}{\bibfnamefont{Z.}~\bibnamefont{Wu}}, \bibnamefont{and}
  \bibinfo{author}{\bibfnamefont{G.~M.} \bibnamefont{Bruun}},
  \bibinfo{journal}{Phys. Rev. A} \textbf{\bibinfo{volume}{94}},
  \bibinfo{pages}{063631} (\bibinfo{year}{2016}).

\bibitem[{\citenamefont{Caracanhas et~al.}(2017)\citenamefont{Caracanhas,
  Schreck, and Smith}}]{Caracanhas2017}
\bibinfo{author}{\bibfnamefont{M.~A.} \bibnamefont{Caracanhas}},
  \bibinfo{author}{\bibfnamefont{F.}~\bibnamefont{Schreck}}, \bibnamefont{and}
  \bibinfo{author}{\bibfnamefont{C.~M.} \bibnamefont{Smith}},
  \bibinfo{journal}{arXiv:1701.04702}.

\bibitem[{\citenamefont{Shankar}(1994)}]{Shankar1994}
\bibinfo{author}{\bibfnamefont{R.}~\bibnamefont{Shankar}},
  \bibinfo{journal}{Rev. Mod. Phys.} \textbf{\bibinfo{volume}{66}},
  \bibinfo{pages}{129} (\bibinfo{year}{1994}).

\bibitem[{\citenamefont{Halboth and Metzner}(2000)}]{Halboth2000}
\bibinfo{author}{\bibfnamefont{C.~J.} \bibnamefont{Halboth}} \bibnamefont{and}
  \bibinfo{author}{\bibfnamefont{W.}~\bibnamefont{Metzner}},
  \bibinfo{journal}{Phys. Rev. B} \textbf{\bibinfo{volume}{61}},
  \bibinfo{pages}{7364} (\bibinfo{year}{2000}).

\bibitem[{\citenamefont{Zanchi and Schulz}(2000)}]{Zanchi2000}
\bibinfo{author}{\bibfnamefont{D.}~\bibnamefont{Zanchi}} \bibnamefont{and}
  \bibinfo{author}{\bibfnamefont{H.~J.} \bibnamefont{Schulz}},
  \bibinfo{journal}{Phys. Rev. B} \textbf{\bibinfo{volume}{61}},
  \bibinfo{pages}{13609} (\bibinfo{year}{2000}).

\bibitem[{\citenamefont{Honerkamp and Salmhofer}(2001)}]{Honerkamp2001}
\bibinfo{author}{\bibfnamefont{C.}~\bibnamefont{Honerkamp}} \bibnamefont{and}
  \bibinfo{author}{\bibfnamefont{M.}~\bibnamefont{Salmhofer}},
  \bibinfo{journal}{Phys. Rev. B} \textbf{\bibinfo{volume}{64}},
  \bibinfo{pages}{184516} (\bibinfo{year}{2001}).

\bibitem[{\citenamefont{Honerkamp}(2001)}]{Honerkamp2001a}
\bibinfo{author}{\bibfnamefont{C.}~\bibnamefont{Honerkamp}},
  \bibinfo{journal}{Eur. Phys. J. B} \textbf{\bibinfo{volume}{21}},
  \bibinfo{pages}{81} (\bibinfo{year}{2001}).

\bibitem[{\citenamefont{Kopietz et~al.}(2010)\citenamefont{Kopietz, Bartosch,
  and Sch{\"{u}}tz}}]{Kopietz2010introduction}
\bibinfo{author}{\bibfnamefont{P.}~\bibnamefont{Kopietz}},
  \bibinfo{author}{\bibfnamefont{L.}~\bibnamefont{Bartosch}}, \bibnamefont{and}
  \bibinfo{author}{\bibfnamefont{F.}~\bibnamefont{Sch{\"{u}}tz}},
  \emph{\bibinfo{title}{{Introduction to the Functional Renormalization
  Group}}}, Lecture Notes in Physics (\bibinfo{publisher}{Springer},
  \bibinfo{address}{Berlin}, \bibinfo{year}{2010}).

\bibitem[{\citenamefont{Metzner et~al.}(2012)\citenamefont{Metzner, Salmhofer,
  Honerkamp, Meden, and Sch{\"{o}}nhammer}}]{Metzner2012}
\bibinfo{author}{\bibfnamefont{W.}~\bibnamefont{Metzner}},
  \bibinfo{author}{\bibfnamefont{M.}~\bibnamefont{Salmhofer}},
  \bibinfo{author}{\bibfnamefont{C.}~\bibnamefont{Honerkamp}},
  \bibinfo{author}{\bibfnamefont{V.}~\bibnamefont{Meden}}, \bibnamefont{and}
  \bibinfo{author}{\bibfnamefont{K.}~\bibnamefont{Sch{\"{o}}nhammer}},
  \bibinfo{journal}{Rev. Mod. Phys.} \textbf{\bibinfo{volume}{84}},
  \bibinfo{pages}{299} (\bibinfo{year}{2012}).

\bibitem[{\citenamefont{Platt et~al.}(2013)\citenamefont{Platt, Hanke, and
  Thomale}}]{Platt2013}
\bibinfo{author}{\bibfnamefont{C.}~\bibnamefont{Platt}},
  \bibinfo{author}{\bibfnamefont{W.}~\bibnamefont{Hanke}}, \bibnamefont{and}
  \bibinfo{author}{\bibfnamefont{R.}~\bibnamefont{Thomale}},
  \bibinfo{journal}{Adv. Phys.} \textbf{\bibinfo{volume}{62}},
  \bibinfo{pages}{453} (\bibinfo{year}{2013}).

\bibitem[{\citenamefont{Ruderman and Kittel}(1954)}]{Ruderman1954}
\bibinfo{author}{\bibfnamefont{M.~A.} \bibnamefont{Ruderman}} \bibnamefont{and}
  \bibinfo{author}{\bibfnamefont{C.}~\bibnamefont{Kittel}},
  \bibinfo{journal}{Phys. Rev.} \textbf{\bibinfo{volume}{96}},
  \bibinfo{pages}{99} (\bibinfo{year}{1954}).

\bibitem[{\citenamefont{Kasuya}(1956)}]{Kasuya1956}
\bibinfo{author}{\bibfnamefont{T.}~\bibnamefont{Kasuya}},
  \bibinfo{journal}{Prog. Theor. Phys.} \textbf{\bibinfo{volume}{16}},
  \bibinfo{pages}{45} (\bibinfo{year}{1956}).

\bibitem[{\citenamefont{Yosida}(1957)}]{Yosida1957}
\bibinfo{author}{\bibfnamefont{K.}~\bibnamefont{Yosida}},
  \bibinfo{journal}{Phys. Rev.} \textbf{\bibinfo{volume}{106}},
  \bibinfo{pages}{893} (\bibinfo{year}{1957}).

\bibitem[{\citenamefont{Illuminati and Albus}(2004)}]{Illuminati2004}
\bibinfo{author}{\bibfnamefont{F.}~\bibnamefont{Illuminati}} \bibnamefont{and}
  \bibinfo{author}{\bibfnamefont{A.}~\bibnamefont{Albus}},
  \bibinfo{journal}{Phys. Rev. Lett.} \textbf{\bibinfo{volume}{93}},
  \bibinfo{pages}{090406} (\bibinfo{year}{2004}).

\bibitem[{\citenamefont{Mathey et~al.}(2006)\citenamefont{Mathey, Tsai, and
  {Castro Neto}}}]{Mathey2006}
\bibinfo{author}{\bibfnamefont{L.}~\bibnamefont{Mathey}},
  \bibinfo{author}{\bibfnamefont{S.-W.} \bibnamefont{Tsai}}, \bibnamefont{and}
  \bibinfo{author}{\bibfnamefont{A.~H.} \bibnamefont{{Castro Neto}}},
  \bibinfo{journal}{Phys. Rev. Lett.} \textbf{\bibinfo{volume}{97}},
  \bibinfo{pages}{030601} (\bibinfo{year}{2006}).

\bibitem[{\citenamefont{Mathey et~al.}(2007)\citenamefont{Mathey, Tsai, and
  {Castro Neto}}}]{Mathey2007}
\bibinfo{author}{\bibfnamefont{L.}~\bibnamefont{Mathey}},
  \bibinfo{author}{\bibfnamefont{S.-W.} \bibnamefont{Tsai}}, \bibnamefont{and}
  \bibinfo{author}{\bibfnamefont{A.~H.} \bibnamefont{{Castro Neto}}},
  \bibinfo{journal}{Phys. Rev. B} \textbf{\bibinfo{volume}{75}},
  \bibinfo{pages}{174516} (\bibinfo{year}{2007}).

\bibitem[{\citenamefont{Lai et~al.}(2014)\citenamefont{Lai, Huang, Campbell,
  and Tsai}}]{Lai2014}
\bibinfo{author}{\bibfnamefont{C.~Y.} \bibnamefont{Lai}},
  \bibinfo{author}{\bibfnamefont{W.~M.} \bibnamefont{Huang}},
  \bibinfo{author}{\bibfnamefont{D.~K.} \bibnamefont{Campbell}},
  \bibnamefont{and} \bibinfo{author}{\bibfnamefont{S.~W.} \bibnamefont{Tsai}},
  \bibinfo{journal}{Phys. Rev. A} \textbf{\bibinfo{volume}{90}},
  \bibinfo{pages}{013610} (\bibinfo{year}{2014}).

\bibitem[{\citenamefont{Pethick and Smith}(2002)}]{pethick2002bose}
\bibinfo{author}{\bibfnamefont{C.~J.} \bibnamefont{Pethick}} \bibnamefont{and}
  \bibinfo{author}{\bibfnamefont{H.}~\bibnamefont{Smith}},
  \emph{\bibinfo{title}{{Bose-Einstein Condensation in Dilute Gases}}}
  (\bibinfo{publisher}{Cambridge University Press},
  \bibinfo{address}{Cambridge}, \bibinfo{year}{2002}).

\bibitem[{\citenamefont{Lewenstein et~al.}(2012)\citenamefont{Lewenstein,
  Sanpera, and Ahufinger}}]{lewenstein2012ultracold}
\bibinfo{author}{\bibfnamefont{M.}~\bibnamefont{Lewenstein}},
  \bibinfo{author}{\bibfnamefont{A.}~\bibnamefont{Sanpera}}, \bibnamefont{and}
  \bibinfo{author}{\bibfnamefont{V.}~\bibnamefont{Ahufinger}},
  \emph{\bibinfo{title}{{Ultracold Atoms in Optical Lattices: Simulating
  Quantum Many-Body Systems}}} (\bibinfo{publisher}{Oxford University Press},
  \bibinfo{address}{Oxford}, \bibinfo{year}{2012}).

\bibitem[{\citenamefont{Blaizot and Ripka}(1986)}]{blaizot1986quantum}
\bibinfo{author}{\bibfnamefont{J.~P.} \bibnamefont{Blaizot}} \bibnamefont{and}
  \bibinfo{author}{\bibfnamefont{G.}~\bibnamefont{Ripka}},
  \emph{\bibinfo{title}{{Quantum Theory of Finite Systems}}}
  (\bibinfo{publisher}{MIT Press}, \bibinfo{address}{Cambridge, MA},
  \bibinfo{year}{1986}).

\bibitem[{\citenamefont{Reiss et~al.}(2007)\citenamefont{Reiss, Rohe, and
  Metzner}}]{Reiss2007}
\bibinfo{author}{\bibfnamefont{J.}~\bibnamefont{Reiss}},
  \bibinfo{author}{\bibfnamefont{D.}~\bibnamefont{Rohe}}, \bibnamefont{and}
  \bibinfo{author}{\bibfnamefont{W.}~\bibnamefont{Metzner}},
  \bibinfo{journal}{Phys. Rev. B} \textbf{\bibinfo{volume}{75}},
  \bibinfo{pages}{075110} (\bibinfo{year}{2007}).

\bibitem[{\citenamefont{Yamase et~al.}(2016)\citenamefont{Yamase, Eberlein, and
  Metzner}}]{Yamase2016}
\bibinfo{author}{\bibfnamefont{H.}~\bibnamefont{Yamase}},
  \bibinfo{author}{\bibfnamefont{A.}~\bibnamefont{Eberlein}}, \bibnamefont{and}
  \bibinfo{author}{\bibfnamefont{W.}~\bibnamefont{Metzner}},
  \bibinfo{journal}{Phys. Rev. Lett.} \textbf{\bibinfo{volume}{116}},
  \bibinfo{pages}{096402} (\bibinfo{year}{2016}).

\bibitem[{\citenamefont{Parsons et~al.}(2016)\citenamefont{Parsons, Mazurenko,
  Chiu, Ji, Greif, and Greiner}}]{Parsons2016}
\bibinfo{author}{\bibfnamefont{M.~F.} \bibnamefont{Parsons}},
  \bibinfo{author}{\bibfnamefont{A.}~\bibnamefont{Mazurenko}},
  \bibinfo{author}{\bibfnamefont{C.~S.} \bibnamefont{Chiu}},
  \bibinfo{author}{\bibfnamefont{G.}~\bibnamefont{Ji}},
  \bibinfo{author}{\bibfnamefont{D.}~\bibnamefont{Greif}}, \bibnamefont{and}
  \bibinfo{author}{\bibfnamefont{M.}~\bibnamefont{Greiner}},
  \bibinfo{journal}{Science} \textbf{\bibinfo{volume}{353}},
  \bibinfo{pages}{1253} (\bibinfo{year}{2016}).

\bibitem[{\citenamefont{Wille et~al.}(2008)\citenamefont{Wille, Spiegelhalder,
  Kerner, Naik, Trenkwalder, Hendl, Schreck, Grimm, Tiecke, Walraven
  et~al.}}]{Wille2008}
\bibinfo{author}{\bibfnamefont{E.}~\bibnamefont{Wille}},
  \bibinfo{author}{\bibfnamefont{F.~M.} \bibnamefont{Spiegelhalder}},
  \bibinfo{author}{\bibfnamefont{G.}~\bibnamefont{Kerner}},
  \bibinfo{author}{\bibfnamefont{D.}~\bibnamefont{Naik}},
  \bibinfo{author}{\bibfnamefont{A.}~\bibnamefont{Trenkwalder}},
  \bibinfo{author}{\bibfnamefont{G.}~\bibnamefont{Hendl}},
  \bibinfo{author}{\bibfnamefont{F.}~\bibnamefont{Schreck}},
  \bibinfo{author}{\bibfnamefont{R.}~\bibnamefont{Grimm}},
  \bibinfo{author}{\bibfnamefont{T.~G.} \bibnamefont{Tiecke}},
  \bibinfo{author}{\bibfnamefont{J.~T.~M.} \bibnamefont{Walraven}}
  \bibnamefont{\textit{et~al.}}, \bibinfo{journal}{Phys. Rev. Lett.}
  \textbf{\bibinfo{volume}{100}}, \bibinfo{pages}{053201}
  (\bibinfo{year}{2008}).

\bibitem[{\citenamefont{Voigt et~al.}(2009)\citenamefont{Voigt, Taglieber,
  Costa, Aoki, Wieser, H{\"{a}}nsch, and Dieckmann}}]{Voigt2009}
\bibinfo{author}{\bibfnamefont{A.~C.} \bibnamefont{Voigt}},
  \bibinfo{author}{\bibfnamefont{M.}~\bibnamefont{Taglieber}},
  \bibinfo{author}{\bibfnamefont{L.}~\bibnamefont{Costa}},
  \bibinfo{author}{\bibfnamefont{T.}~\bibnamefont{Aoki}},
  \bibinfo{author}{\bibfnamefont{W.}~\bibnamefont{Wieser}},
  \bibinfo{author}{\bibfnamefont{T.~W.} \bibnamefont{H{\"{a}}nsch}},
  \bibnamefont{and}
  \bibinfo{author}{\bibfnamefont{K.}~\bibnamefont{Dieckmann}},
  \bibinfo{journal}{Phys. Rev. Lett.} \textbf{\bibinfo{volume}{102}},
  \bibinfo{pages}{020405} (\bibinfo{year}{2009}).

\bibitem[{\citenamefont{Spiegelhalder et~al.}(2009)\citenamefont{Spiegelhalder,
  Trenkwalder, Naik, Hendl, Schreck, and Grimm}}]{Spiegelhalder2009}
\bibinfo{author}{\bibfnamefont{F.~M.} \bibnamefont{Spiegelhalder}},
  \bibinfo{author}{\bibfnamefont{A.}~\bibnamefont{Trenkwalder}},
  \bibinfo{author}{\bibfnamefont{D.}~\bibnamefont{Naik}},
  \bibinfo{author}{\bibfnamefont{G.}~\bibnamefont{Hendl}},
  \bibinfo{author}{\bibfnamefont{F.}~\bibnamefont{Schreck}}, \bibnamefont{and}
  \bibinfo{author}{\bibfnamefont{R.}~\bibnamefont{Grimm}},
  \bibinfo{journal}{Phys. Rev. Lett.} \textbf{\bibinfo{volume}{103}},
  \bibinfo{pages}{223203} (\bibinfo{year}{2009}).

\bibitem[{\citenamefont{Tiecke et~al.}(2010)\citenamefont{Tiecke, Goosen,
  Ludewig, Gensemer, Kraft, Kokkelmans, and Walraven}}]{Tiecke2010}
\bibinfo{author}{\bibfnamefont{T.~G.} \bibnamefont{Tiecke}},
  \bibinfo{author}{\bibfnamefont{M.~R.} \bibnamefont{Goosen}},
  \bibinfo{author}{\bibfnamefont{A.}~\bibnamefont{Ludewig}},
  \bibinfo{author}{\bibfnamefont{S.~D.} \bibnamefont{Gensemer}},
  \bibinfo{author}{\bibfnamefont{S.}~\bibnamefont{Kraft}},
  \bibinfo{author}{\bibfnamefont{S.~J. J. M.~F.} \bibnamefont{Kokkelmans}},
  \bibnamefont{and} \bibinfo{author}{\bibfnamefont{J.~T.~M.}
  \bibnamefont{Walraven}}, \bibinfo{journal}{Phys. Rev. Lett.}
  \textbf{\bibinfo{volume}{104}}, \bibinfo{pages}{053202}
  (\bibinfo{year}{2010}).

\bibitem[{\citenamefont{Spiegelhalder et~al.}(2010)\citenamefont{Spiegelhalder,
  Trenkwalder, Naik, Kerner, Wille, Hendl, Schreck, and
  Grimm}}]{Spiegelhalder2010}
\bibinfo{author}{\bibfnamefont{F.~M.} \bibnamefont{Spiegelhalder}},
  \bibinfo{author}{\bibfnamefont{A.}~\bibnamefont{Trenkwalder}},
  \bibinfo{author}{\bibfnamefont{D.}~\bibnamefont{Naik}},
  \bibinfo{author}{\bibfnamefont{G.}~\bibnamefont{Kerner}},
  \bibinfo{author}{\bibfnamefont{E.}~\bibnamefont{Wille}},
  \bibinfo{author}{\bibfnamefont{G.}~\bibnamefont{Hendl}},
  \bibinfo{author}{\bibfnamefont{F.}~\bibnamefont{Schreck}}, \bibnamefont{and}
  \bibinfo{author}{\bibfnamefont{R.}~\bibnamefont{Grimm}},
  \bibinfo{journal}{Phys. Rev. A} \textbf{\bibinfo{volume}{81}},
  \bibinfo{pages}{043637} (\bibinfo{year}{2010}).

\bibitem[{\citenamefont{Park et~al.}(2012)\citenamefont{Park, Wu, Santiago,
  Tiecke, Will, Ahmadi, and Zwierlein}}]{Park2012}
\bibinfo{author}{\bibfnamefont{J.~W.}~\bibnamefont{Park}},
  \bibinfo{author}{\bibfnamefont{C.~H.} \bibnamefont{Wu}},
  \bibinfo{author}{\bibfnamefont{I.}~\bibnamefont{Santiago}},
  \bibinfo{author}{\bibfnamefont{T.~G.} \bibnamefont{Tiecke}},
  \bibinfo{author}{\bibfnamefont{S.}~\bibnamefont{Will}},
  \bibinfo{author}{\bibfnamefont{P.}~\bibnamefont{Ahmadi}}, \bibnamefont{and}
  \bibinfo{author}{\bibfnamefont{M.~W.} \bibnamefont{Zwierlein}},
  \bibinfo{journal}{Phys. Rev. A} \textbf{\bibinfo{volume}{85}},
  \bibinfo{pages}{051602} (\bibinfo{year}{2012}).

\bibitem[{\citenamefont{Ospelkaus
  et~al.}(2006{\natexlab{a}})\citenamefont{Ospelkaus, Ospelkaus, Sengstock, and
  Bongs}}]{Ospelkaus2006}
\bibinfo{author}{\bibfnamefont{C.}~\bibnamefont{Ospelkaus}},
  \bibinfo{author}{\bibfnamefont{S.}~\bibnamefont{Ospelkaus}},
  \bibinfo{author}{\bibfnamefont{K.}~\bibnamefont{Sengstock}},
  \bibnamefont{and} \bibinfo{author}{\bibfnamefont{K.}~\bibnamefont{Bongs}},
  \bibinfo{journal}{Phys. Rev. Lett.} \textbf{\bibinfo{volume}{96}},
  \bibinfo{pages}{020401} (\bibinfo{year}{2006}{\natexlab{a}}).

\bibitem[{\citenamefont{Ospelkaus
  et~al.}(2006{\natexlab{b}})\citenamefont{Ospelkaus, Ospelkaus, Humbert,
  Sengstock, and Bongs}}]{Ospelkaus2006a}
\bibinfo{author}{\bibfnamefont{S.}~\bibnamefont{Ospelkaus}},
  \bibinfo{author}{\bibfnamefont{C.}~\bibnamefont{Ospelkaus}},
  \bibinfo{author}{\bibfnamefont{L.}~\bibnamefont{Humbert}},
  \bibinfo{author}{\bibfnamefont{K.}~\bibnamefont{Sengstock}},
  \bibnamefont{and} \bibinfo{author}{\bibfnamefont{K.}~\bibnamefont{Bongs}},
  \bibinfo{journal}{Phys. Rev. Lett.} \textbf{\bibinfo{volume}{97}},
  \bibinfo{pages}{120403} (\bibinfo{year}{2006}{\natexlab{b}}).

\bibitem[{\citenamefont{Best et~al.}(2009)\citenamefont{Best, Will, Schneider,
  Hackerm{\"{u}}ller, van Oosten, Bloch, and L{\"{u}}hmann}}]{Best2009}
\bibinfo{author}{\bibfnamefont{T.}~\bibnamefont{Best}},
  \bibinfo{author}{\bibfnamefont{S.}~\bibnamefont{Will}},
  \bibinfo{author}{\bibfnamefont{U.}~\bibnamefont{Schneider}},
  \bibinfo{author}{\bibfnamefont{L.}~\bibnamefont{Hackerm{\"{u}}ller}},
  \bibinfo{author}{\bibfnamefont{D.}~\bibnamefont{van Oosten}},
  \bibinfo{author}{\bibfnamefont{I.}~\bibnamefont{Bloch}}, \bibnamefont{and}
  \bibinfo{author}{\bibfnamefont{D.~S.} \bibnamefont{L{\"{u}}hmann}},
  \bibinfo{journal}{Phys. Rev. Lett.} \textbf{\bibinfo{volume}{102}},
  \bibinfo{pages}{030408} (\bibinfo{year}{2009}).

\bibitem[{\citenamefont{Silber et~al.}(2005)\citenamefont{Silber,
  G{\"{u}}nther, Marzok, Deh, Courteille, and Zimmermann}}]{Silber2005}
\bibinfo{author}{\bibfnamefont{C.}~\bibnamefont{Silber}},
  \bibinfo{author}{\bibfnamefont{S.}~\bibnamefont{G{\"{u}}nther}},
  \bibinfo{author}{\bibfnamefont{C.}~\bibnamefont{Marzok}},
  \bibinfo{author}{\bibfnamefont{B.}~\bibnamefont{Deh}},
  \bibinfo{author}{\bibfnamefont{P.~W.} \bibnamefont{Courteille}},
  \bibnamefont{and}
  \bibinfo{author}{\bibfnamefont{C.}~\bibnamefont{Zimmermann}},
  \bibinfo{journal}{Phys. Rev. Lett.} \textbf{\bibinfo{volume}{95}},
  \bibinfo{pages}{170408} (\bibinfo{year}{2005}).

\bibitem[{\citenamefont{Ivanov et~al.}(2011)\citenamefont{Ivanov, Khramov,
  Hansen, Dowd, M{\"{u}}nchow, Jamison, and Gupta}}]{Ivanov2011}
\bibinfo{author}{\bibfnamefont{V.~V.} \bibnamefont{Ivanov}},
  \bibinfo{author}{\bibfnamefont{A.}~\bibnamefont{Khramov}},
  \bibinfo{author}{\bibfnamefont{A.~H.} \bibnamefont{Hansen}},
  \bibinfo{author}{\bibfnamefont{W.~H.} \bibnamefont{Dowd}},
  \bibinfo{author}{\bibfnamefont{F.}~\bibnamefont{M{\"{u}}nchow}},
  \bibinfo{author}{\bibfnamefont{A.~O.} \bibnamefont{Jamison}},
  \bibnamefont{and} \bibinfo{author}{\bibfnamefont{S.}~\bibnamefont{Gupta}},
  \bibinfo{journal}{Phys. Rev. Lett.} \textbf{\bibinfo{volume}{106}},
  \bibinfo{pages}{153201} (\bibinfo{year}{2011}).

\bibitem[{\citenamefont{Hara et~al.}(2011)\citenamefont{Hara, Takasu, Yamaoka,
  Doyle, and Takahashi}}]{Hara2011}
\bibinfo{author}{\bibfnamefont{H.}~\bibnamefont{Hara}},
  \bibinfo{author}{\bibfnamefont{Y.}~\bibnamefont{Takasu}},
  \bibinfo{author}{\bibfnamefont{Y.}~\bibnamefont{Yamaoka}},
  \bibinfo{author}{\bibfnamefont{J.~M.} \bibnamefont{Doyle}}, \bibnamefont{and}
  \bibinfo{author}{\bibfnamefont{Y.}~\bibnamefont{Takahashi}},
  \bibinfo{journal}{Phys. Rev. Lett.} \textbf{\bibinfo{volume}{106}},
  \bibinfo{pages}{205304} (\bibinfo{year}{2011}).

\bibitem[{\citenamefont{Sugawa et~al.}(2010)\citenamefont{Sugawa, Inaba, Taie,
  Yamazaki, Yamashita, and Takahashi}}]{Sugawa2010}
\bibinfo{author}{\bibfnamefont{S.}~\bibnamefont{Sugawa}},
  \bibinfo{author}{\bibfnamefont{K.}~\bibnamefont{Inaba}},
  \bibinfo{author}{\bibfnamefont{S.}~\bibnamefont{Taie}},
  \bibinfo{author}{\bibfnamefont{R.}~\bibnamefont{Yamazaki}},
  \bibinfo{author}{\bibfnamefont{M.}~\bibnamefont{Yamashita}},
  \bibnamefont{and}
  \bibinfo{author}{\bibfnamefont{Y.}~\bibnamefont{Takahashi}},
  \bibinfo{journal}{Nat. Phys.} \textbf{\bibinfo{volume}{7}},
  \bibinfo{pages}{642} (\bibinfo{year}{2011}).

\bibitem[{\citenamefont{Carusotto and Castin}(2005)}]{Carusotto2005}
\bibinfo{author}{\bibfnamefont{I.}~\bibnamefont{Carusotto}} \bibnamefont{and}
  \bibinfo{author}{\bibfnamefont{Y.}~\bibnamefont{Castin}},
  \bibinfo{journal}{Phys. Rev. Lett.} \textbf{\bibinfo{volume}{94}},
  \bibinfo{pages}{223202} (\bibinfo{year}{2005}).

\bibitem[{\citenamefont{Dao et~al.}(2007)\citenamefont{Dao, Georges, Dalibard,
  Salomon, and Carusotto}}]{Dao2007}
\bibinfo{author}{\bibfnamefont{T.~L.} \bibnamefont{Dao}},
  \bibinfo{author}{\bibfnamefont{A.}~\bibnamefont{Georges}},
  \bibinfo{author}{\bibfnamefont{J.}~\bibnamefont{Dalibard}},
  \bibinfo{author}{\bibfnamefont{C.}~\bibnamefont{Salomon}}, \bibnamefont{and}
  \bibinfo{author}{\bibfnamefont{I.}~\bibnamefont{Carusotto}},
  \bibinfo{journal}{Phys. Rev. Lett.} \textbf{\bibinfo{volume}{98}},
  \bibinfo{pages}{240402} (\bibinfo{year}{2007}).

\bibitem[{\citenamefont{Gritsev et~al.}(2008)\citenamefont{Gritsev, Demler, and
  Polkovnikov}}]{Gritsev2008}
\bibinfo{author}{\bibfnamefont{V.}~\bibnamefont{Gritsev}},
  \bibinfo{author}{\bibfnamefont{E.}~\bibnamefont{Demler}}, \bibnamefont{and}
  \bibinfo{author}{\bibfnamefont{A.}~\bibnamefont{Polkovnikov}},
  \bibinfo{journal}{Phys. Rev. A} \textbf{\bibinfo{volume}{78}},
  \bibinfo{pages}{063624} (\bibinfo{year}{2008}).

\bibitem[{\citenamefont{Pekker et~al.}(2009)\citenamefont{Pekker, Sensarma, and
  Demler}}]{Pekker2009}
\bibinfo{author}{\bibfnamefont{D.}~\bibnamefont{Pekker}},
  \bibinfo{author}{\bibfnamefont{R.}~\bibnamefont{Sensarma}}, \bibnamefont{and}
  \bibinfo{author}{\bibfnamefont{E.}~\bibnamefont{Demler}},
  \bibinfo{journal}{arXiv:0906.0931}.

\bibitem[{\citenamefont{Altman et~al.}(2004)\citenamefont{Altman, Demler, and
  Lukin}}]{Altman2004}
\bibinfo{author}{\bibfnamefont{E.}~\bibnamefont{Altman}},
  \bibinfo{author}{\bibfnamefont{E.}~\bibnamefont{Demler}}, \bibnamefont{and}
  \bibinfo{author}{\bibfnamefont{M.~D.} \bibnamefont{Lukin}},
  \bibinfo{journal}{Phys. Rev. A} \textbf{\bibinfo{volume}{70}},
  \bibinfo{pages}{013603} (\bibinfo{year}{2004}).

\bibitem[{\citenamefont{Kitagawa et~al.}(2011)\citenamefont{Kitagawa, Aspect,
  Greiner, and Demler}}]{Kitagawa2011}
\bibinfo{author}{\bibfnamefont{T.}~\bibnamefont{Kitagawa}},
  \bibinfo{author}{\bibfnamefont{A.}~\bibnamefont{Aspect}},
  \bibinfo{author}{\bibfnamefont{M.}~\bibnamefont{Greiner}}, \bibnamefont{and}
  \bibinfo{author}{\bibfnamefont{E.}~\bibnamefont{Demler}},
  \bibinfo{journal}{Phys. Rev. Lett.} \textbf{\bibinfo{volume}{106}},
  \bibinfo{pages}{115302} (\bibinfo{year}{2011}).

\end{thebibliography}
\end{document}